\documentclass[12pt]{article}

\usepackage{graphicx}
\usepackage{amsmath,amssymb}
\usepackage{bm}
\usepackage{graphicx, color}
\usepackage{wrapfig}
\usepackage{cite}

\usepackage{geometry}
\renewcommand{\Im}{\mathrm{Im}}
\renewcommand{\det}{\mathrm{det}}
\newcommand{\tr}{\mathrm{Tr}}

\begin{document}
\title{
\begin{flushright}
\ \\*[-80pt]
\begin{minipage}{0.2\linewidth}
\normalsize
HUPD 1902 \\*[50pt]
\end{minipage}
\end{flushright}
{\Large \bf
Sign of CP Violating Phase in Quarks and Leptons
\\*[20pt]}}

\author{
	\centerline{
		~Yusuke~Shimizu$^{1,}$\footnote{E-mail address: yu-shimizu@hiroshima-u.ac.jp},
		~Kenta~Takagi$^{1,}$\footnote{E-mail address: takagi-kenta@hiroshima-u.ac.jp},} \\*[10pt]
	\centerline{Shunya~Takahashi$^{1,}$\footnote{E-mail address: s-takahashi@hiroshima-u.ac.jp},
		~Morimitsu~Tanimoto$^{2,}$\footnote{E-mail address: tanimoto@muse.sc.niigata-u.ac.jp}}
	\\*[30pt]
	\centerline{
		\begin{minipage}{\linewidth}
			\begin{center}
				$^1${\it \normalsize
					Graduate School of Science, Hiroshima University, Higashi-Hiroshima 739-8526, Japan} \\*[5pt]
				$^2${\it \normalsize
					Department of Physics, Niigata University, Niigata 950-2181, Japan}
			\end{center}
	\end{minipage}}
\\*[50pt]}
\date{
\centerline{\small \bf Abstract}
\begin{minipage}{0.9\linewidth}
	\medskip
	\medskip
	\medskip
	\medskip
\small
We discuss the relation between the CP violation of the quark mixing and that of the lepton mixing by investigating a CP violating observable, the Jarlskog invariant, as well as the CP violating Dirac phase.
The down-type quark mass matrix with three zeros is given in terms of the minimal number of parameters, while the up-type quark mass matrix is diagonal.
These quark mass matrices leading to the successful CKM mixing angles and CP violation are embedded in both the Pati--Salam and SU(5) models.
The leptonic Jarlskog invariant $J_{CP}^l$ (as well as CP violating Dirac phase) is examined for two cases:
the neutrino mass matrix is diagonal or non-diagonal,
where no additional CP violating phase is introduced apart from the Majorana phases.
In the case of the diagonal neutrino mass matrix, the favorable sign of the leptonic CP violation is obtained, however, the magnitude of $J_{CP}^l$ is at most ${\cal O}(10^{-4})$, which is too small compared with the expected value from the observation $-0.02$.
In the case of the non-diagonal neutrino mass matrix where the tri-bimaximal mixing pattern is taken, we obtain the successful $J_{CP}^l$ up to its sign.
\end{minipage}
}

\begin{titlepage}
\maketitle
\thispagestyle{empty}
\end{titlepage}

\section{Introduction}

The standard model is well established by the discovery of the Higgs boson.
However, the flavor theory of the quark and lepton mass matrices is still unknown.
Therefore, we do not know a principle to determine the quark and lepton mass matrices.
A long time ago, Weinberg \cite{Weinberg:1977hb} considered a mass matrix for the down-type quarks in the basis of the up-type quark mass matrix being diagonal.
He assumed a vanishing (1,1) element in a $2\times2$ matrix and the matrix to be symmetric in order to reduce the number of free parameters.
Then, the number of free parameters is reduced to only two and hence he succeeded to predict the Cabibbo angle to be $\sqrt{m_d/m_s}$, which is very successful and called as the Gatto, Sartori, Tonin relation \cite{Gatto:1968ss}.
Fritzsch extended the above approach to the three family case \cite{Fritzsch:1977vd,Fritzsch:1979zq} where four zeros were set in both down-type and up-type quark mass matrices and these matrices were assumed to be Hermitian.
Then, there were eight parameters against the ten observed data.
However, it was ruled out by the observed Cabibbo--Kobayashi--Maskawa (CKM) matrix element $V_{cb}$
\footnote{More general study of four zeros  was given in the nearest neighbor interactions \cite{Branco:1988iq}. }.
Ramond, Roberts and Ross also presented a systematic work with four or five zeros in the symmetric or Hermitian quark mass matrix \cite{Ramond:1993kv}.
Their textures are also disfavored under the precise experimental data at present, because four or five zeros is too tight to reproduce the ten observed data \cite{Fritzsch:2002ga}.

One of authors and Yanagida proposed quark mass matrices in the standpoint of  ``Occam's Razor'' approach \cite{Tanimoto:2016rqy}
\footnote{
	The ``Occam's Razor'' approach was at first introduced in the neutrino sector by \cite{Harigaya:2012bw}, and its phenomenological discussions appeared in \cite{Kaneta:2016gbq,Shimizu:2016eih}.}
, where the minimum number of parameters was taken for the successful CKM mixing angles and CP violation without assuming the symmetric or Hermitian mass matrix of down-type quarks.
Three zeros were imposed in the down-type quark mass matrix, and the up-type quark mass matrix was assumed to be diagonal.
Therefore, the down-type quark mass matrix was given with six complex parameters.
Among them, five phases are removed by the phase redefinition of the three right-handed and three left-handed down-type quark fields.
After the field-phase rotation, there remain six real parameters and one phase which is the source of the CP violation.
These seven parameters are the minimal number to reproduce the seven observed data: the three down-type quark masses and the four CKM parameters.
It is emphasized that the three zeros are maximal zeros to keep one CP violating phase in the down-type quark mass matrix.
This framework reproduces  observed CKM mixing angles and the CP violating phase  succesfully  \cite{Tanimoto:2016rqy}.
Indeed, it is found that thirteen textures of down-type quark mass matrix are completely consistent with the present experimental data of down-type quark masses and CKM parameters.

The phase in the down-type quark mass matrix is directly related to the CP violating phase in the CKM matrix.
If the down-type quark mass matrix is related to the charged lepton mass matrix in the quark-lepton unification,
the CP violating phase of the quark sector also appears in the charged lepton mass matrix.
Then, the CP violating observable of the quark sector is correlated with that of the lepton sector.
Linking the leptonic CP violation to the quark unitarity triangle is an attractive work \cite{Tanimoto:2015hqa} in order to develop the flavor physics of quarks and leptons
since the recent experiments of neutrino oscillations  strongly indicate the CP violation \cite{Abe:2018wpn,Adamson:2017gxd}.

We consider the unification of quarks and leptons in the Pati--Salam model and the SU(5) model of grand unified theory (GUT).
The charged lepton mass matrix is given by the down-type quark mass matrix with the Clebsch--Gordan (CG) coefficients
which have been investigated comprehensively in the renormalizable or non-renormalizable operators of dimensions 4, 5, and 6 \cite{Antusch:2009gu}.
Such CG coefficients are necessary to reproduce the proper mass ratios of quarks and leptons.
The CG coefficient for the SU(5) model is the well-known Georgi--Jarlskog factor: $-3$,
if the renormalizable Yukawa couplings come only from renormalizable operators and the Higgs sits in a $\bf \overline {45}$ of SU(5) \cite{Georgi:1979df}.
For the case of non-renormalizable operators, systematic studies have been presented focusing on the neutrino mixing angles in the Pati--Salam model and the SU(5) model \cite{Antusch:2011qg,Marzocca:2011dh}.

In this work, we embed above successful quark mass matrices in the Pati--Salam model
\footnote{
	The Pati--Salam model with different textures of up- and down-type quarks cannot be embedded into an SO(10) GUT
	since up- and down-type components of the 16-dimensional representation of SO(10) transform in the same manner.
	However, we can consider the Pati--Salam gauge group emerged directly from string theory \cite{King:2014iia}.}
and the SU(5) model by assuming that one single operator dominates each matrix element
\footnote{It is remarked that the
 neutrino mixing angles and the Dirac phase were obtained successfully for an asymmetric mass matrix of the down-type quarks in SU(5) and SO(10) \cite{Rahat:2018sgs}.}.
We discuss correlations between the Jarlskog invariant (as well as the CP violating Dirac phase) of quarks and that of leptons by choosing relevant CG coefficients comprehensively.
Our investigation is taken place for two cases:
the neutrino mass matrix is diagonal or non-diagonal,
where no additional CP violating phase is introduced apart from the Majorana phases.

In section 2, we summarize viable down-type quark mass matrices with three zeros.
In section 3, we calculate the Jarlskog invariant and the CP violating Dirac phase of both quarks and leptons in the Pati--Salam model and the SU(5) model of GUT.
Section 4 is devoted to the summary of our work.
In Appendix A, we show redundancy of our textures.
In Appendix B, we show how to calculate the CKM mixing angles and the CP violation explicitly.
In Appendix C, we present the charged lepton mass matrices in the Pati--Salam model and the SU(5) model.
In Appendix D, we present the relevant formulae of the CP violation.


\section{Three texture zeros for down-type quarks }

In the standpoint of the ``Occam's Razor'' approach \cite{Tanimoto:2016rqy},
the down-type quark mass matrix has been investigated by putting zeros at several elements of the mass matrix in the basis of  the diagonal up-type quark mass matrix.
Then, the number of free parameters of the mass matrix is reduced.
It is found that the three texture zeros provide the minimum number of parameters needed for the successful CKM mixing angles and CP violation.

Let us define the Lagrangian for the Yukawa couplings in the quark sector as follows:
\begin{align}
	\mathcal{L}_Y=-y_{\alpha\beta}^u \bar Q_{L\alpha} u_{R\beta} \tilde h
	-y_{\alpha\beta}^d \bar Q_{L\alpha} d_{R\beta} h \ ,
\label{lagrangian}
\end{align}
where $Q_{L\alpha}$,  $u_{R\beta}$,  $d_{R\beta}$ and $h$ denote the left-handed quark doublets, the right-handed up-type quark singlet,
the right-handed down-type quark singlet, and  the Higgs doublet, respectively.
The indices $\alpha$ and $\beta$ denote flavors.
The quark mass matrices are given as $m_{\alpha\beta}=y_{\alpha\beta} v_H$ with $v_H=174.1$ GeV.
In order to reproduce the observed quark masses and the CKM matrix with the minimal number of parameters,
we take a diagonal basis for the up-type quark mass matrix:
\begin{equation}
	M_U= {\rm diag}\,\{m_u, m_c, m_t\}\,.
\label{up}
\end{equation}

For the down-type quark mass matrix, we impose the three texture zeros.
Then, the texture of the down-type quark mass matrix $M_D$ is given with six complex parameters.
The five phases can be removed by the phase rotation of the three right-handed and three left-handed down-type quark fields.
Therefore, there remain six real parameters and one CP violating phase,
which are the minimal number to reproduce the observed data of down-type quark masses and the CKM parameters.

Now, we discuss textures for the down-type quark mass matrix.
Let us start with taking $(3,3)$, $(2,3)$, $(1,2)$ elements of $M_D$ to be non-vanishing values to reproduce the observed bottom quark mass and the CKM mixing angles, $V_{us}$ and $V_{cb}$.
Then, we have $_6C_3=20$ textures with three zeros for the down-type quark mass matrix.
For our convenience, we classify them in two categories, (A) and (B)
\footnote{
	The classification of categories A and B is only for our convenience
	because the freedom of unitary transformations of the right-handed down-type quarks spoil this classification as discussed in Appendix A. }.
We have $_5 C_2=10$ textures with a non-vanishing $(2,2)$ element in (A) and 10 textures with a vanishing  $(2,2)$ element in (B).

In the category (A), the following six textures are consistent with the present experimental data.
Those six down-type quark mass matrices are parametrized after removing five phases by the phase rotation of quark fields as follows:
\begin{align}
	&M_D^{(1)}=
	\begin{pmatrix}
	0 & a_D & 0 \\
	a'_D & b_D \ e^{-i\phi}& c_D \\
	0 & c'_D & d_D
	\end{pmatrix}_{LR} , ~
	M_{D}^{(2)}=
	\begin{pmatrix}
	a'_D & a_D & 0 \\
	0 & b_D \ e^{-i\phi}& c_D \\
	0 & c'_D & d_D
	\end{pmatrix}_{LR} , ~
	M_{D}^{(3)}=
	\begin{pmatrix}
	0 & a_D & 0 \\
	0 & b_D \ e^{-i\phi}& c_D \\
	a'_D & c'_D & d_D
	\end{pmatrix}_{LR}  , \nonumber\\
	& M_{D}^{(4)}=
	\begin{pmatrix}
	0 & a_D & c'_D \\
	a'_D &  b_D \ e^{-i\phi}& c_D \\
	0 & 0 & d_D
	\end{pmatrix}_{LR}  , ~
	M_{D}^{(5)}=
	\begin{pmatrix}
	a'_D & a_D  & c'_D \\
	0 & b_D \ e^{-i\phi}& c_D \\
	0 & 0 & d_D
	\end{pmatrix}_{LR}  , ~
	M_{D}^{(6)}=
	\begin{pmatrix}
	0 & a_D & c'_D \\
	0 & b_D \ e^{-i\phi}& c_D \\
	a'_D & 0 & d_D
	\end{pmatrix}_{LR} ,
\label{downmassmatrixA}
\end{align}
where $a_D, a'_D, b_D, c_D, c'_D$ and $d_D$ are real parameters and $\phi$ is a CP violating phase.
It should be stressed that our matrices are not symmetric at all.
The CP violating phase $\phi$ is put in the $(2,2)$ entry.
In the category (B), the $(2,2)$ element is zero.
The following seven textures are also consistent with the present experimental data.
After removing five phases by the phase rotation of quark fields, those seven down-type quark mass matrices are parametrized as
\begin{align}
	&M_D^{(11)}=
	\begin{pmatrix}
	a'_D &a_D \ e^{-i\phi} & b_D \\
	0 & 0 \ & c_D \\
	0 & c'_D & d_D
	\end{pmatrix}_{LR}  , ~
	M_{D}^{(12)}=
	\begin{pmatrix}
	0 & a_D \ e^{-i\phi} & b_D \\
	a'_D& 0& c_D \\
	0 & c'_D & d_D
	\end{pmatrix}_{LR}  , ~
	M_{D}^{(13)}=
	\begin{pmatrix}
	0 & a_D \ e^{-i\phi} & b_D \\
	0 & 0 & c_D \\
	a'_D & c'_D & d_D
	\end{pmatrix}_{LR}  , \nonumber\\
	&M_{D}^{(14)}=
	\begin{pmatrix}
	a_D\ e^{i\phi} & a'_D  & c'_D \\
	b_D & 0 & c_D \\
	0 & 0 & d_D
	\end{pmatrix}_{LR}  , ~
	M_{D}^{(15)}=
	\begin{pmatrix}
	 a_D \ e^{-i\phi}& a'_D & b_D \\
	0 & 0 & c_D \\
	c'_D & 0 & d_D
	\end{pmatrix}_{LR} , \nonumber\\
	&M_{D}^{(16)}=
	\begin{pmatrix}
	0  & a_D & b_D \\
	a'_D & 0 & c_D \ e^{-i\phi}\\
	c'_D & 0 & d_D
	\end{pmatrix}_{LR} ,  ~
	M_{D}^{(17)}=
	\begin{pmatrix}
	a_D & a'_D & 0 \\
	b_D & 0& c_D \ e^{i\phi}\\
	c'_D & 0 & d_D
	\end{pmatrix}_{LR}  .
\label{downmassmatrixB}
\end{align}

We comment on the freedoms of unitary transformations of the right-handed quarks.
Since the CKM matrix is the flavor mixing among the left-handed quarks,
some textures in Eqs.(\ref{downmassmatrixA}) and (\ref{downmassmatrixB}) are equivalent each other due to the freedom of unitary transformations of the right-handed quarks.
We show the redundancy among them in Appendix A.

\begin{table}[hbtp]
	\begin{center}
		\begin{tabular}{l|c c c c c}
			\hline
			& $|V_{us}|$ & $|V_{cb}|$ &$|V_{ub}|$& $\delta_{CP}^q$& $j_{CP}^q$
			\rule[14pt]{1pt}{0pt} \\
			\hline
			$M_{D}^{(1)}$,$M_{D}^{(2)}$, $M_{D}^{(3)}$,$M_{D}^{(16)}$, $M_{D}^{(17)}$
			& $\frac{a_D b_D}{m_s^2} \left |\sin\frac{\phi}{2}\right |$  &
			$\frac{\sqrt{2}c_D}{m_b}\left |\cos\frac{\phi}{2}\right |$&  $\frac{a_D c'_D}{m_b^2}$& $\frac{1}{2}(\pi-\phi)$&
			$a^2_D b_D c_D c'_D d_D \sin\phi$\\
			$M_{D}^{(4)}$,$M_{D}^{(5)}$,$M_{D}^{(6)}$,$M_{D}^{(14)}$ & $\frac{a_D b_D}{m_s^2}$& $\frac{c_D}{m_b}$& $\frac{c'_D}{m_b}$& $\phi$&$a_D b_D c_D c'_D d_D^2 \sin\phi$\\
			$M_{D}^{(11)}$,$M_{D}^{(12)}$,$M_{D}^{(13)}$,$M_{D}^{(15)}$& $\frac{a_D c_D}{m_s^2}\frac{c'_D}{m_b}$& $\frac{c_D}{m_b}$& $\frac{b_D}{m_b}$
			& $\pi-\phi$&$a_D b_D c^2_D c'_D d_D \sin\phi$\\
			\hline
		\end{tabular}
		\caption{The predicted CKM matrix elements, the CP phase and the CP violating measure $j_{CP}^q$ \cite{Tanimoto:2016rqy},
			where $|V_{ij}|$ and $\delta_{CP}^q$ are given in the leading order while $j_{CP}^q$ is the exact one.}
		\label{tab1}
	\end{center}
\end{table}

The CKM matrix elements $|V_{ij}|$ and the CP violating phase $\delta_{CP}^q$ are given approximately as shown in Table \ref{tab1} \cite{Tanimoto:2016rqy},
where $\delta_{CP}^q$ is defined by PDG convention \cite{Tanabashi:2018oca}.
There is another CP violating observable, the Jarlskog invariant $J_{CP}^q$ \cite{Jarlskog:1985ht}.
It is derived from the following relation
\cite{Bernabeu:1986fc,Gronau:1986xb,Branco:1999fs,Branco:2011zb}
  as presented in Appendix D:
\begin{align}
	& {\rm Tr}([H_U, H_D]^3) = 6 i J^q_{CP} \Delta_u \Delta_d \ ,
\label{Jcp}
\end{align}
where 
\begin{align}
	 H_U \equiv M_U M_U^\dagger  \ , \qquad\qquad H_D \equiv M_D M_D^\dagger ~,
\end{align}
and
\begin{align}
	\Delta_u
	\equiv
	(m_u^2 - m_t^2)(m_u^2 - m_c^2)(m_c^2 - m_t^2)<0, \qquad
	\Delta_d
	\equiv
	(m_d^2 - m_b^2)(m_d^2 - m_s^2)(m_s^2 - m_b^2)<0.
\end{align}
The predicted $J_{CP}^q$ is exactly expressed in terms of the parameters of the mass matrix elements
and it is consistent with the observed value \cite{Tanabashi:2018oca}
after fixing parameters of the matrix elements:
\begin{eqnarray}
	\qquad\qquad J_{CP}^q=(3.18\pm 0.15)\times 10^{-5}~ .
	\label{JCP}
\end{eqnarray}
In Table 1, we present $j^q_{CP}$ instead of $J^q_{CP}$ which is related as follows:
\begin{align}
	j^q_{CP}\equiv -\Delta_d~  J^q_{CP}  \ .
\end{align}
We summarize  three CKM matrix elements $V_{ij}$, the  CP violating phase $\delta_{CP}^q$ and $j_{CP}^q$ in Table 1 for $M_D^{(k)}$ $(k=1\,\mbox{--}\,6, 11\,\mbox{--}\,17)$.

We can obtain a parameter region in which the observed CKM mixing angles and the CP violating phase are reproduced
by inputting the three observed down-type quark masses and four CKM parameters (see Appendix B).
In Ref. \cite{Tanimoto:2016rqy}, quark masses and CKM parameters
are taken at the electroweak (EW) scale \cite{Antusch:2013jca} because
Xing and Zhao \cite{Xing:2015sva} found that texture zeros of the quark mass matrix are essentially stable against the evolution apart from the magnitudes of matrix elements.
The allowed region of parameters are presented in Table 4 of Appendix B.

In the present paper, we use quark masses and CKM parameters at the GUT scale since we discuss the quark-lepton unification.
Indeed, we adopt quark masses and CKM parameters at the GUT scale in minimal supersymmetric standard model (MSSM)
\cite{Antusch:2013jca,Bjorkeroth:2015ora},
which is also discussed in Appendix B.
For the case of $M_D^{(1)}$, we obtain a parameter region as follows:
\begin{align}
	&a_D =(4.2-6.4)\times 10^{-3}, \quad   a'_D =(2.2-5.0)\times 10^{-3}, \quad
	b_D=(24-48)\times 10^{-3},\nonumber\\
	&c_D=(18-46)\times 10^{-3}, \quad ~~  c'_D=0.61-0.89,
	\quad\qquad~~~ d_D=0.82-1.04, ~\qquad\phi=(23-61)^\circ,
\label{bench0}
\end{align}
in GeV unit apart from $\phi$,
which are reduced by approximately $1/3$ compared with values of at the EW scale apart from $\phi$ as seen in Table 4.
In our numerical discussions of the next section, we take the central values of these parameters:
\begin{align}
	&a_D =5.3\times 10^{-3}, \qquad\quad   a'_D =3.6\times 10^{-3}, \qquad\quad
	b_D=36\times 10^{-3}, \nonumber\\
	&c_D=32\times 10^{-3},\qquad\quad~  c'_D=0.75,
	\qquad\qquad\quad ~~~ d_D=0.93,
	\qquad\qquad \phi=42^\circ,
\label{bench}
\end{align}
as a benchmark.


\section{CP violation of leptons}

We have viable 13 mass matrices of down-type quarks
and discuss the flavor mixing in the lepton sector especially for the CP violation in the Pati--Salam GUT and the SU(5) GUT.
The charged lepton mass matrix is related to the down-type quark mass matrix with CG coefficients in these GUT models.
The CG coefficients are necessary to reproduce the relevant mass ratios of quarks and leptons.
Possible CG coefficients have been obtained in the renormalizable or non-renormalizable operators of dimensions 4, 5, and 6 \cite{Antusch:2009gu}.
In the case of the Pati--Salam symmetry, possible CG coefficients for dimensions 4, 5, and 6 are
\begin{equation}
	{\rm dimension~4} : (1,  -3) ~ , \qquad
	{\rm dimension~5} : (1,  -3, 9) ~ ,\qquad
	{\rm dimension~6} : (0, \frac{3}{4}, 1, 2, -3)~ .
\label{CG-Pati}
\end{equation}
For SU(5), we have possible CG coefficients:
\begin{equation}
	{\rm dimension~4} : (1,  -3) ~ , \qquad
	{\rm dimension~5} : (-\frac{1}{2},1, \pm\frac{3}{2},  -3, \frac{9}{2}, 6, 9, -18) ~ .
\label{CG-SU5}
\end{equation}

Let us discuss the charged lepton mass matrix corresponding to $M_D^{(1)}$ as a representative.
In the Pati--Salam model, it is
\begin{align}
	M_E^{(1)}=
	\begin{pmatrix}
	0 & a_E & 0 \\
	a'_E & b_E \ e^{-i\phi}& c_E \\
	0 & c'_E & d_E
	\end{pmatrix}_{LR} .
\label{charged-1-Pati}
\end{align}
In the SU(5) model, the charged lepton mass matrix is the transpose of 
$M_D^{(1)}$:
\begin{align}
	M_E^{(1)}=
	\begin{pmatrix}
	0 & a'_E & 0 \\
	a_E & b_E \ e^{-i\phi}& c'_E \\
	0 & c_E & d_E
	\end{pmatrix}_{LR} . ~
\label{charged-1-SU5}
\end{align}
We assume that one single operator dominates each matrix element.
Then, the charged lepton mass matrix elements are given in terms of  the down-type quark mass matrix elements and CG coefficients as follows:
\begin{equation}
	a_E= C_a a_D,  \quad  a'_E= C_{a'} a'_D, \quad  b_E= C_b b_D, \quad
	c_E= C_c c_D, \quad  c'_E= C_{c'} c'_D, \quad  d_E= C_d d_D,
\label{E-element}
\end{equation}
where $C_a, C_{a'}, C_b, C_{c}, C_{c'}$, and $C_d$ are possible CG coefficients in Eq.(\ref{CG-Pati}) for the Pati--Salam model or in Eq.(\ref{CG-SU5}) for the SU(5) model.
The phase $\phi$ is common to the phase in the down-type quark mass matrix of Eq.(\ref{downmassmatrixA}).

\subsection{Flavor mixing only from charged leptons}

In this subsection, we discuss the case where the lepton flavor mixing comes from only the charged lepton mass matrix.
The neutrinos are supposed to be Majorana particles and their mass matrix is to be diagonal.
They have three different mass eigenvalues $(m_1, m_2, m_3)$ with Majorana phases.
We examine carefully if this model works or not by estimating the magnitude of the leptonic CP violating measure $J_{CP}^l$.

\subsubsection{Pati--Salam model with diagonal neutrino mass matrix}
The charged lepton mass matrix $M_E^{(1)}$ for the Pati--Salam model is given in Eq.(\ref{charged-1-Pati}).
Then we have
\begin{equation}
	M_E^{(1)} M_E^{(1)\dagger}=
	\begin{pmatrix}
	a^2_E & a_E b_E\  e^{i\phi} & a_E c'_E \\
	a_E b_E\  e^{-i\phi} &a'^2_E+b_E^2+c_E^2& b_E c'_E\  e^{-i\phi}+
	c_E d_E\\
	a_E c'_E &  b_E c'_E\  e^{i\phi}+c_E d_E & c'^2_E+d^2_E
	\end{pmatrix} \ .
\label{MMdagger1}
\end{equation}
The left-handed mixing matrix of the charged leptons
$U_E$ is given as a diagonalizing matrix of $M_E^{(1)} M_E^{(1)\dagger}$:
\begin{align}
	U_E  M_E^{(1)} M_E^{(1)\dagger} U_E^\dagger=
	{\rm diag}\{m_e^2,m_\mu^2, m_\tau^2 \} ~ .
\end{align}
We show $U_E^\dagger$  in the leading order:
\begin{align}
	U_E^\dagger
	\simeq
	\begin{pmatrix}
	X_e & \frac{a_E b_E}{m_\mu^2}e^{i(\pi+\phi)/2}\sin\frac{\phi}{2}Y_\mu & \frac{a_E c'_E}{m_\tau^2}Z_\tau\\[4pt]
	\frac{a_E}{2b_E\sin\frac{\phi}{2}}e^{i(\pi-\phi)/2}X_e & Y_\mu & \frac{b_E d_E}{m_\tau^2}\cos\frac{\phi}{2}e^{-i\phi/2}Z_\tau\\[4pt]
	\frac{a_E}{2d_E\sin\frac{\phi}{2}}e^{-i(\pi+\phi)/2}X_e &
	-\frac{c_E}{d_E}\cos\frac{\phi}{2} e^{i\phi/2}Y_\mu & Z_\tau
	\end{pmatrix}~,
\end{align}
where $X_{e}$, $Y_{\mu}$, and  $Z_{\tau}$ are determined by the normalization condition.
If the off-diagonal elements are much smaller than $1$,
$X_{e}$, $Y_{\mu}$, and $Z_{\tau}$ are $1$ in a good approximation.
However, the present case is non-trivial due to the observed large mixing angles in the lepton sector.

The Pontecorvo--Maki--Nakagawa--Sakata (PMNS) matrix is defined as $U_{\mathrm{PMNS}}=U_E U_\nu^\dagger$ where $U_\nu$ is the left-handed mixing matrix for the neutrinos.
Since the neutrino mass matrix is supposed to be diagonal, the PMNS matrix is
\begin{align}
	U_{\mathrm{PMNS}} = U_E \simeq
	\begin{pmatrix}
	X_e & \frac{a_E}{2b_E\sin\frac{\phi}{2}}e^{-i(\pi-\phi)/2}X_e &
	\frac{a_E}{2d_E\sin\frac{\phi}{2}}e^{i(\pi+\phi)/2}X_e \\[4pt]
	\frac{a_E b_E}{m_\mu^2}e^{-i(\pi+\phi)/2}\sin\frac{\phi}{2}Y_\mu
	 & Y_\mu &-\frac{c_E}{d_E}\cos\frac{\phi}{2} e^{-i\phi/2}Y_\mu  \\[4pt]
	\frac{a_E c'_E}{m_\tau^2}Z_\tau &
	\frac{b_E d_E}{m_\tau^2}\cos\frac{\phi}{2}e^{i\phi/2}Z_\tau
	& Z_\tau
	\end{pmatrix}~.
\end{align}
After rephasing the mixing matrix,  we obtain the PMNS matrix in the PDG convention \cite{Tanabashi:2018oca} as follows:
\begin{align}
	U_{\mathrm{PMNS}}
	\simeq
	\begin{pmatrix}
	X_e & \frac{a_E}{2b_E\sin\frac{\phi}{2}}X_e &
	\frac{a_E}{2d_E\sin\frac{\phi}{2}}e^{i\phi/2}X_e \\[4pt]
	-\frac{a_E b_E}{m_\mu^2}\sin\frac{\phi}{2}Y_\mu
	& Y_\mu &\frac{c_E}{d_E}\cos\frac{\phi}{2} Y_\mu  \\[4pt]
	\frac{a_E c'_E}{m_\tau^2}e^{i\pi/2}Z_\tau &
	-\frac{b_E d_E}{m_\tau^2}Z_\tau
	& Z_\tau
	\end{pmatrix}~.
\end{align}
The CP violating Dirac phase is given as $\delta_{CP}^l\equiv -\arg [U_{\rm PMNS}(1,3)]\simeq -\phi/2$.
If we compare $\delta_{CP}^l$ with $\delta_{CP}^q\simeq (\pi-\phi)/2$ in Table 1,
we have a sum rule of the CP violating phase between the quark and lepton sector:
 \begin{align}
 \delta_{CP}^l \simeq \delta_{CP}^q -\frac{\pi}{2} ~ ,
 \end{align}
which gives $\delta_{CP}^l\simeq -20^\circ$ by inputting $\delta_{CP}^q \simeq 70^\circ$ \cite{Tanabashi:2018oca}.
Indeed, the numerical calculation gives  $\delta_{CP}^l\simeq -20.4^\circ$
as shown later.
The predicted minus sign is favored.
However, we should examine whether this PMNS matrix can leads to two large mixing angles while reproducing the observed mass hierarchy of the charged leptons by taking the relevant CG coefficients
in Eq.(\ref{CG-Pati}).

In order to test the consistency of  this PMNS matrix  with the large flavor mixing angles,
we calculate the magnitude of the leptonic CP violating measure $J_{CP}^l$
\cite{Krastev:1988yu} .  
It  is expected to be much larger than that of quarks $J_{CP}^q$
due to two large flavor mixing angles of the lepton sector.
Indeed, the present best fit value of $J_{CP}^l$ in the global analysis NuFIT \cite{NuFIT} is given at the EW scale as:
\begin{align}
	J_{CP}^l \simeq -2\times 10^{-2}~ ,
\label{JCPl}
\end{align}
which is  $10^3$ times larger than the observed one of quarks in Eq.(\ref{JCP}).


As presented in Appendix D,
the leptonic CP violating measure is obtained \cite{Bernabeu:1986fc,Gronau:1986xb,Branco:1999fs,Branco:2011zb} as
\footnote{The neutrino mass term is defined as  $-\frac{1}{2}\overline{\nu_L}M_\nu (\nu_L)^c$
as given in Appendix D.
 Our  $M_\nu$ corresponds to $\mathbf{m}_\nu^*$  in \cite{Branco:2011zb}. }
\begin{align}
	\mathrm{Tr}([H_\nu,H_E]^3)	=
	-6iJ_{CP}^l ~\Delta_{\nu} ~\Delta_{e} ~ ,
\label{eq:Tr_nu_E}
\end{align}
where $H_\nu = M_\nu M_\nu^\dagger$ and $H_E = M_E M_E^\dagger$, and
\begin{align}
	\Delta_\nu
	\equiv
	(m_1^2 - m_3^2)(m_1^2 - m_2^2)(m_2^2 - m_3^2)<0~,
	\qquad \Delta_e
	\equiv
	(m_e^2 - m_\tau^2)(m_e^2 - m_\mu^2)(m_\mu^2 - m_\tau^2)<0~.
\end{align}
Therefore, we can obtain $J_{CP}^l$ in terms of the charged lepton matrix elements
from $H_E$ of Eq.(\ref{MMdagger1}) and the diagonal $H_\nu$.
With a result of $\mathrm{Tr}([H_\nu,H_E]^3)$ for $M_D^{(1)}$ as presented in Table 2, we obtain
\begin{align}
	J_{CP}^l =~ \frac{1}{\Delta_e}~  a_E^2 b_E c_E c'_E d_E  \sin\phi ~ .
\end{align}
Since these parameters of matrix elements are related to the down-type quark parameters by CG coefficients as seen in Eq.(\ref{E-element}),
we can express $J_{CP}^l$ in terms of CG coefficients  in Eq.(\ref{CG-Pati}) as:
\begin{align}
	J_{CP}^l =~ \frac{1}{\Delta_e}~  a_D^2 b_D c_D c'_D d_D ~ C_a^2 C_b C_c C_{c'} C_{d} \sin\phi ~ ,
\label{JCPlepton}
\end{align}
where $\Delta_e$ is also given by the mass matrix elements approximately:
\begin{align}
	\Delta_e\simeq -m_\tau^4 m_\mu^2~, \qquad
	 m_\tau^2\simeq c'^2_E+d_E^2~,\qquad
	 m_\tau^2m_\mu^2\simeq c^2_E c'^2_E+b^2_E d^2_E -2b_Ec_E c'_E d_E\cos\phi ~ ,
\end{align}
as seen in Appendix B.
There is a simple constraint for CG coefficients through the determinant of the mass matrices:
\begin{align}
	\det (M_D^{(1)}M_D^{(1)\dagger})=m_d^2 m_s^2 m_b^2 = a_D^2 a'^2_D d^2_D~,
	\qquad
	\det (M_E^{(1)}M_E^{(1)\dagger})
	=m_e^2 m_\mu^2 m_\tau^2 = a_E^2 a'^2_E d^2_E~.
\end{align}
By taking a ratio of the two equations, we have
\begin{align}
	C_a^2~ C_{a'}^2 ~ C_d^2=
	\frac{m_e^2 ~m_\mu^2~ m_\tau^2 }{m_d^2~ m_s^2 ~m_b^2 } =1.7\,\mbox{--}\,7.3~,
\label{detratio}
\end{align}
where   Yukawa couplings of quarks and leptons are taken  at the GUT scale
of MSSM  \cite{Bjorkeroth:2015ora} (see Appendix B).

There is another constraint for $C_d$ and $C_{c'}$.
As seen in Appendix B, a relation $y_b\simeq y_\tau$ is given by the observations at the GUT scale.
Since we have $m_\tau^2\simeq d_E^2+ c'^2_E$ and $m_b^2\simeq d_D^2+ c'^2_D$, we should choose
$C_d=  1$ and $C_{c'}= 1$ from Eq.(\ref{CG-Pati}).

The last constraint comes from the following ratio (see also Appendix B),
\begin{align}
	\frac{m_e^2 m_\mu^2+ m_\mu^2 m_\tau^2 +m_e^2 m_\tau^2 }
	{m_d^2 m_s^2+ m_s^2 m_b^2 +m_d^2 m_b^2}
	\simeq \frac{c^2_E c'^2_E+b^2_E d^2_E -2b_Ec_E c'_E d_E\cos\phi }
	{c^2_D c'^2_D+b^2_D d^2_D -2b_Dc_D c'_D d_D\cos\phi} ~ ,
\end{align}
which is $15\,\mbox{--}\,26$ at the GUT scale.

The CG coefficients should be restricted so as to reproduce $C_a^2 C_{a'}^2  C_d^2 = {\cal O}(1)$ and $C_a^2 C_b C_c C_{c'} C_{d} \gg 1$ as seen in Eqs.(\ref{detratio}) and (\ref{JCPlepton}).
Finally, we choose
\begin{eqnarray}
	C_a=2 , \quad  C_{a'}=1,
	\quad  C_b=-3,  \quad  C_c=-3, \quad  C_{c'}=1 , \quad  C_d=1 ~,
\end{eqnarray}
which leads to
\begin{align}
	J_{CP}^l=-8.1\times 10^{-5}~, \qquad \frac{m_e^2 ~m_\mu^2~ m_\tau^2 }{m_d^2~ m_s^2 ~m_b^2 } =4 ~ ,
	\qquad \frac{m_e^2 m_\mu^2+ m_\mu^2 m_\tau^2 +m_e^2 m_\tau^2 }
	{m_d^2 m_s^2+ m_s^2 m_b^2 +m_d^2 m_b^2} =8.5 ~,
\end{align}
where the last mass ratio $8.5$ is rather small compared to the observed value $15\,\mbox{--}\,26$.
It is noticed that the calculated $J_{CP}^l$ is two order smaller than the expected value in Eq.(\ref{JCPl}) although the predicted minus sign is favored.

Thus, the Pati--Salam model with $M_E^{(1)}$ in the case where the neutrino mass matrix is diagonal leads to the inconsistent prediction of $J_{CP}^l$ with the observed value in Eq.\eqref{JCPl}.
Indeed, the failure of the prediction for the magnitude of $J_{CP}^l$ is due to the wrong prediction of the lepton large mixing angles:
\begin{align}
\sin^2 \theta_{12}^{\rm PMNS}\simeq 0.021 , \qquad  \sin^2\theta_{23}^{\rm PMNS} \simeq 0.012, \qquad \sin \theta_{13}^{\rm PMNS}\simeq 0.015, \qquad
\delta_{CP}^l\simeq  -20.4^\circ ,
\end{align}
where we take the PDG convention for the mixing angles and the phase \cite{Tanabashi:2018oca}.
In order to overcome too small $J_{CP}^l$ in the Pati--Salam model, we consider a non-diagonal neutrino mass matrix in the next subsection.

For other cases of the charged lepton mass matrices presented in Appendix C-1, we can also examine the magnitude of $J_{CP}^l$ while reproducing the proper mass ratio of down-type quarks and charged leptons.
We have presented the summary of $\mathrm{Tr}([H_\nu,H_E]^3)$ and $\det (H_E)$ in  Table 2.
It is remarked that the sign of  $J_{CP}^l$ is opposite to the quark CP violation $J_{CP}^q$ as far as the sign of CG coefficients are positive.
However, the magnitudes of  $J_{CP}^l$ for all cases of textures are inconsistent with the expected value of observations in the Pati--Salam model.

\begin{table}[htb]
	\begin{center}
		\begin{tabular}{cccc}
			\hline
 			$M_D$
 			&  $\tr\left([H_\nu,H_E]^3\right)_{\rm Pati-Salam}$
 			&  $\tr\left([H_\nu,H_E]^3\right)_{\rm SU(5)}$
 			& $\det(H_E)$\\[2pt]
 			\hline
 			$M_D^{(1)}$
 			& $-6ia_E^2b_Ec_Ec_E'd_E\Delta_\nu\sin\phi$
 			& $-6ia_E'^2b_Ec_Ec_E'd_E\Delta_\nu\sin\phi$
 			& $a_E^2a_E'^2d_E^2$\\[4pt]
 			$M_D^{(2)}$
 			& $-6ia_E^2b_Ec_Ec_E'd_E\Delta_\nu\sin\phi$
 			& 0
 			& $a_E'^2(c_E^2c_E'^2+b_E^2d_E^2-2b_Ec_Ec_E'd_E\cos\phi)$\\[4pt]
 			$M_D^{(3)}$
 			& $-6ia_E^2b_Ec_Ec_E'd_E\Delta_\nu\sin\phi$
 			& $6ia_E'^2b_Ec_Ec_E'd_E\Delta_\nu\sin\phi$
 			& $a_E^2a_E'^2c_E^2$\\[4pt]
 			$M_D^{(4)}$
 			& $-6ia_Eb_Ec_Ec_E'd_E^2\Delta_\nu\sin\phi$
 			& $-6ia_Ea_E'^2b_Ec_Ec_E'\Delta_\nu\sin\phi$
 			& $a_E^2a_E'^2d_E^2$\\[4pt]
 			$M_D^{(5)}$
 			& $-6ia_Eb_Ec_Ec_E'd_E^2\Delta_\nu\sin\phi$
 			& $6ia_Ea_E'^2b_Ec_Ec_E'\Delta_\nu\sin\phi$
 			& $a_E'^2b_E^2d_E^2$\\[4pt]
 			$M_D^{(6)}$
 			& $-6ia_Eb_Ec_Ec_E'd_E^2\Delta_\nu\sin\phi$
 			& 0
 			& $a_E'^2(a_E^2c_E^2 + b_E^2c_E'^2 -2a_Eb_Ec_Ec_E'\cos\phi)$\\[4pt]
 			$M_D^{(11)}$
 			& $-6ia_Eb_Ec_E^2c_E'd_E\Delta_\nu\sin\phi$
 			& $-6ia_Ea_E'^2b_Ec_E'd_E\Delta_\nu\sin\phi$
 			& $a_E'^2c_E^2c_E'^2$\\[4pt]
 			$M_D^{(12)}$
 			& $-6ia_Eb_Ec_E^2c_E'd_E\Delta_\nu\sin\phi$
 			& 0
 			& $a_E'^2(b_E^2c_E'^2 + a_E^2d_E^2 -2a_Eb_Ec_E'd_E\cos\phi)$\\[4pt]
 			$M_D^{(13)}$
 			& $-6ia_Eb_Ec_E^2c_E'd_E\Delta_\nu\sin\phi$
 			&  $6ia_Ea_E'^2b_Ec_E'd_E\Delta_\nu\sin\phi$
 			& $a_E^2a_E'^2c_E^2$\\[4pt]
 			$M_D^{(14)}$
 			& $-6ia_Eb_Ec_Ec_E'd_E^2\Delta_\nu\sin\phi$
 			& $-6ia_Ea_E'^2b_Ec_Ec_E'\Delta_\nu\sin\phi$
 			& $a_E'^2b_E^2d_E^2$\\[4pt]
 			$M_D^{(15)}$
 			& $-6ia_Eb_Ec_E^2c_E'd_E\Delta_\nu\sin\phi$
 			& $6ia_Ea_E'^2b_Ec_E'd_E\Delta_\nu\sin\phi$
 			& $a_E'^2c_E^2c_E'^2$\\[4pt]
 			$M_D^{(16)}$
 			& $-6ia_E'b_E^2c_Ec_E'd_E\Delta_\nu\sin\phi$
 			& 0
 			& $a_E^2(c_E^2c_E'^2 + a_E'^2d_E^2 -2a_E'c_Ec_E'd_E\cos\phi)$\\[4pt]
 			$M_D^{(17)}$
 			& $-6ia_E^2b_Ec_Ec_E'd_E\Delta_\nu\sin\phi$
 			& 0
 			& $a_E'^2(c_E^2c_E'^2 + b_E^2d_E^2 -2b_Ec_Ec_E'd_E\cos\phi)$\\[4pt]
 			\hline
 		\end{tabular}
 	\caption{Summary of $\tr\left([H_\nu,H_E]^3\right)$ and $\det(H_E)$ for the Pati--Salam model and the SU(5) model with the diagonal neutrino mass matrix. Note that $\tr\left([H_\nu,H_E]^3\right)=
 		-6i\Delta_\nu \Delta_e J_{\mathrm{CP}}^l$.
 		}
	\end{center}
\end{table}


\subsubsection{SU(5) model with diagonal neutrino mass matrix}


The charged lepton mass matrix $M_E^{(1)}$ for the SU(5) model is given in Eq.(\ref{charged-1-SU5}).
Then, we have
\begin{align}
	M_E^{(1)}M_E^{(1)\dagger}=
	\begin{pmatrix}
	a'^2_E & a'_E b_E e^{i\phi} & a'_E c_E\\
	a'_E b_E e^{-i\phi} & a_E^2+b_E^2+c'^2_E & b_Ec_E e^{-i\phi} +
	c'_Ed_E\\
	a'_Ec_E & b_Ec_E e^{i\phi} +c'_Ed_E & c^2_E+d_E^2
	\end{pmatrix}~.
\end{align}
We obtain the left-handed mixing matrix of the charged lepton $U_E$ by diagonalizing $M_E^{(1)}M_E^{(1)\dagger}$:
\begin{align}
	U_E  M_E^{(1)} M_E^{(1)\dagger} U_E^\dagger=
	{\rm diag}\{m_e^2,m_\mu^2, m_\tau^2 \} ~ .
\end{align}
The leading order of $U_E^\dagger$ is given as:
\begin{align}
	U_E^\dagger
	\simeq
	\begin{pmatrix}
	X_e & \frac{2a'_E b_E}{m_\mu^2}e^{i(\pi+\phi)/2}\sin\frac{\phi}{2}Y_\mu & \frac{a'_E b_E}{c'^2_E-m_\tau^2}e^{i(\pi+\phi)}Z_\tau\\[4pt]
	\frac{a'_E}{2b_E\sin\frac{\phi}{2}}e^{i(\pi-\phi)/2}X_e & Y_\mu & -\frac{c'_E d_E}{c'^2_E-m_\tau^2}Z_\tau\\[4pt]
	\frac{a'_E c'_E(b_E d_E e^{-i\phi}-c_E c'_E)}{\left|b_E d_E-c_E c'_Ee^{i\phi}\right|^2}X_e &
	-\frac{c'_E}{d_E}Y_\mu & Z_\tau
	\end{pmatrix}~,
\end{align}
where $X_{e}$, $Y_{\mu}$, and $Z_{\tau}$ are determined by the normalization condition.

The PMNS matrix is given as:
\begin{align}
	U_{\mathrm{PMNS}}=U_E\simeq
	\begin{pmatrix}
	X_e & \frac{a_E'}{2b_E\sin\frac{\phi}{2}}e^{-i(\pi-\phi)/2}X_e & \frac{a_E'c_E'(b_Ed_Ee^{i\phi}-c_Ec_E')}{\left|b_Ed_E-c_Ec_E'e^{i\phi}\right|^2}X_e\\[4pt]
	\frac{2a_E'b_E}{m_\mu^2}e^{-i(\pi+\phi)/2}\sin\frac{\phi}{2}Y_\mu & Y_\mu & -\frac{c_E'}{d_E}Y_\mu\\[4pt]
	\frac{a_E'b_E}{c_E'^2-m_\tau^2}e^{-i(\pi+\phi)}Z_\tau & -\frac{c_E'd_E}{c_E'^2-m_\tau^2}Z_\tau & Z_\tau
	\end{pmatrix}~.
\end{align}
Finally, we obtain the PMNS matrix in the following form after rephasing it:
\begin{align}
	U_{\mathrm{PMNS}}
	\simeq
	\begin{pmatrix}
	X_e
	&\frac{a_E'}{2b_E\sin\frac{\phi}{2}}X_e
	&-\frac{a_E'c_E'(b_Ed_Ee^{i\phi}-c_Ec_E')}{\left|b_Ed_E-c_Ec_E'e^{i\phi}\right|^2}e^{i(\pi-\phi)/2}X_e\\[4pt]
	-\frac{2a_E'b_E}{m_\mu^2}\sin\frac{\phi}{2}Y_\mu
	& Y_\mu
	& \frac{c_E'}{d_E}Y_\mu\\[4pt]
	\frac{a_E'b_E}{c_E'^2-m_\tau^2}e^{-i(\pi+\phi)/2}Z_\tau
	& \frac{c_E'd_E}{c_E'^2-m_\tau^2}Z_\tau
	& Z_\tau
	\end{pmatrix}~.
\end{align}
The CP violating Dirac phase is given as
$\delta_{CP}^l\equiv -\arg [U_{\rm PMNS}(1,3)]= 0^\circ$
if  $b_Ed_E= c_Ec_E'$ is put in the numerator of the (1,3) entry.
Indeed, $\delta_{CP}^l\simeq -1.7^\circ$ is obtained  as shown later.

We also calculate $J_{CP}^l$ directly by Eq.(\ref{eq:Tr_nu_E}) as:
\begin{align}
	J_{CP}^l = \frac{1}{\Delta_e} a'^2_E b_E c_E c'_E d_E  \sin\phi ~ ,
\end{align}
which is expressed also in terms of the down-type quark parameters and CG coefficients as follows:
\begin{align}
	J_{CP}^l = \frac{1}{\Delta_e} a'^2_D b_D c_D c'_D d_D C_{a'}^2 C_b C_c C_{c'} C_{d} \sin\phi ~ ,
\label{JCPleptonSU5}
\end{align}
where CG coefficients are given in Eq.(\ref{CG-SU5}).

By a similar investigation of the mass ratios in the Pati--Salam model, the choice of  CG coefficients are restricted.
We find
\begin{eqnarray}
	C_a=1 , \quad  C_{a'}=\frac{3}{2},
	\quad  C_b=\frac{9}{2}, \quad  C_c=6, \quad  C_{c'}=1 , \quad  C_d=1 ~,
\end{eqnarray}
which leads to
\begin{align}
	 J_{CP}^l=- 2.5\times 10^{-5}, \qquad
	\frac{m_e^2 ~m_\mu^2~ m_\tau^2 }{m_d^2~ m_s^2 ~m_b^2 }
	=2.3  , \qquad \frac{m_e^2 m_\mu^2+ m_\mu^2 m_\tau^2 +m_e^2 m_\tau^2 }
	{m_d^2 m_s^2+ m_s^2 m_b^2 +m_d^2 m_b^2}= 20.6 ,
\end{align}
where the mass ratios are consistent with observed ones.
The predicted $J_{CP}^l$ is three order smaller than the expected value in Eq.(\ref{JCPl}) although the predicted minus sign is favored.
Moreover, the PMNS mixing angle $\sin^2\theta_{12}^{\rm PMNS}$ is  very small compared with the observed one \cite{NuFIT}:

\begin{align}
\sin^2 \theta_{12}^{\rm PMNS}\simeq 0.0022 , \qquad  \sin^2\theta_{23}^{\rm PMNS} \simeq 0.39, \qquad \sin \theta_{13}^{\rm PMNS}\simeq 0.038, \qquad
\delta_{CP}^l\simeq  -1.7^\circ .
\end{align}
Thus, the SU(5) model with $M_E^{(1)}$ in the case where the neutrino mass matrix is diagonal also leads to the inconsistent prediction of $J_{CP}^l$ with the expected value in Eq.\eqref{JCPl}.
It is remarked that the sign of  $J_{CP}^l$ is opposite to the quark CP violation $J_{CP}^q$ as far as the sign of CG coefficients are positive.

For other cases of the charged lepton mass matrices presented in Appendix C-2,
we can examine the magnitude of $J_{CP}^l$ while reproducing the proper mass ratio of the down-type quarks and the charged leptons.
It is remarked that $J_{CP}^l$ vanishes for $M_E^{(2)}$, $M_E^{(6)}$, $M_E^{(12)}$, $M_E^{(16)}$, and $M_E^{(17)}$ due to zero textures .
For other non-zero cases, the magnitudes of $J_{CP}^l$ are also inconsistent with the expected value from observations in the SU(5) model.
We have presented the summary of $\mathrm{Tr}([H_\nu,H_E]^3)$ and $\det (H_E)$ in  Table 2.

\subsection{Flavor mixing from both charged leptons and neutrinos}
The magnitude of  $J_{CP}^l$ is two order smaller or less compared with the observed value Eq.(\ref{JCPl}) in the both Pati--Salam model and the SU(5) model
if the neutrino mass matrix is diagonal.
In other words, the large two mixing angles of the PMNS matrix are not reproduced in the framework of the Pati--Salam model and the SU(5) model only by the charged lepton mass matrix with three zeros.

 On the other hand, we know interesting ideas to relate the CKM matrix and the PMNS matrix, "the quark-lepton complementarity" \cite{Minakata:2004xt} and "Cabibbo haze" \cite{Datta:2005ci}. 
   In both approaches, the large lepton mixing angles come from the neutrino
   sector, but the link of the CKM matrix appears through the charged lepton sector.
  These approaches motivate us to consider  a non-diagonal neutrino mass matrix in order to obtain the two large mixing angles.
  
This situation can be derived in the seesaw mechanism with the non-diagonal right-handed Majorana neutrino mass matrix while  the  Dirac neutrino mass matrix is still diagonal.  Then, the new Dirac CP phase appears as well as the Majorana phases in general.  However, if the (1-3) flavor mixing angle $\theta_{13}$ of the neutrino sector
vanishes or is negligibly small, this new Dirac CP phase 
does not contribute to the PMNS matrix. 
Therefore, we can study the correlation between the CP violating Dirac phases in the quarks and leptons
because the only CP violating phase $\phi$ is still common in both the quark and lepton sector.
Majorana phases do not affect our analysis.
Let us consider the case of vanishing  $\theta_{13}$.
We can parametrize an orthogonal matrix which diagonalizes the neutrino mass matrix as
\begin{align}
	U_\nu
	=
	\begin{pmatrix}
	\cos\theta_{12} & \sin\theta_{12} & 0\\
	-\cos\theta_{23}\sin\theta_{12} & \cos\theta_{12}\cos\theta_{23} & -\sin\theta_{23}\\
	-\sin\theta_{12}\sin\theta_{23} & \cos\theta_{12}\sin\theta_{23} & \cos\theta_{23}
	\end{pmatrix} P ~,
\label{neutrinomixing}
\end{align}
where  $P$ is a diagonal $3\times 3$ Majorana phase matrix.
The neutrino mass matrix is given as:
\begin{align}
	M_\nu
	=
	U_\nu P^*
	\begin{pmatrix}
	m_1 & 0 & 0\\
	0 & m_2 & 0\\
	0 & 0 & m_3
	\end{pmatrix}
	P^* U_\nu^T~.
\label{neutrinomassmatrix}
\end{align}
It is easy to find that the Majorana phase matrix $P$ disappears in $H_\nu\equiv M_\nu M_\nu^\dagger$.


\subsubsection{Pati--Salam model with non-diagonal neutrino mass matrix}
Let us start to discuss the case of the Pati--Salam model.
The charged lepton mass matrix is $M_E^{(1)}$ in Eq.(\ref{charged-1-Pati}) while the neutrino mass matrix is in Eq.(\ref{neutrinomassmatrix}).
We can calculate $J_{CP}^l$ directly by use of the formula in Eq.(\ref{eq:Tr_nu_E}).
We obtain
\begin{align}
	\mathrm{Tr}([H_\nu , H_E^{(1)}]^3)
	&=-6iJ_{CP}^l ~\Delta_{\nu} ~\Delta_{e}
	 \nonumber\\
	&=
	-\frac{3}{4}i \Delta_{\nu}
	a_Eb_E ~
	\left[
	4a_Ec_E'\cos(2\theta_{12})
	\left\{
	2c_Ed_E\cos(2\theta_{23})
	-(a_E'^2 + c_E^2 - d_E^2)\sin(2\theta_{23})
	\right\}
	\right.\nonumber\\
	&\phantom{=}\left.
	+\sin(2\theta_{12})
	\left\{
	-
	\left[
	(a_E'^2 +c_E^2)(a_E^2-4c_E'^2)
	-(a_E^2 +a_E'^2 +b_E^2-c_E^2-c_E'^2)d_E^2+d_E^4
	\right]\cos\theta_{23}
	\right.\right.\nonumber\\
	&\phantom{=}\left.\left.
	+\left\{
	a_E^2(a_E'^2 + c_E^2)
	-(a_E^2 +a_E'^2 +b_E^2 +3c_E^2-c_E'^2)d_E^2
	+d_E^4
	\right\}\cos(3\theta_{23})\right.\right.\nonumber\\
	&\phantom{=}\left.\left.
	+c_Ed_E
	(-2a_E^2+a_E'^2+b_E^2+c_E^2+3c_E'^2+d_E^2)\sin\theta_{23}\right.\right.\nonumber\\
	&\phantom{=}\left.\left.
	-8b_Ec_E'd_E(c_E\cos\theta_{23}+d_E\sin\theta_{23})\cos^2\theta_{23}\cos\phi\right.\right.\nonumber\\
	&\phantom{=}\left.\left.
	+c_Ed_E(2a_E^2 + a_E'^2+b_E^2+c_E^2-c_E'^2-3d_E^2)\sin(3\theta_{23})
	\right\}\right]\sin\phi~.
\end{align}
We write $J_{CP}^l$ in terms of the leading order by taking account of the
magnitudes of parameters  in Eq.(\ref{bench}):
\begin{align}
	J_{CP}^l
	&\simeq
	\frac{1}{8\Delta_e}
	a_Eb_E
	\left[
	4a_Ec_E'\cos(2\theta_{12})
	\left\{
	2c_Ed_E\cos(2\theta_{23})
	+d_E^2\sin(2\theta_{23})
	\right\}
	\right.\nonumber\\
	&\phantom{=}\left.
	+\sin(2\theta_{12})\left\{
	-d_E^2(c_E'^2 + d_E^2)\left[\cos\theta_{23}-\cos(3\theta_{23})\right]
	+c_Ed_E(3c_E'^2+d_E^2)\sin\theta_{23}
	\right.\right.\nonumber\\
	&\phantom{=}\left.\left.
	-8b_Ec_E'd_E(c_E\cos\theta_{23}+d_E\sin\theta_{23})\cos^2\theta_{23}\cos\phi
	\right.\right.\nonumber\\
	&\phantom{=}\left.\left.
	-c_Ed_E(c_E'^2 + 3d_E^2)\sin(3\theta_{23})
	\right\}
	\right]\sin\phi~.
\end{align}
If we suppose $\mathcal{O}(\sin\theta_{12})\sim\mathcal{O}(\cos\theta_{12})$ and
$\mathcal{O}(\sin\theta_{23})\sim\mathcal{O}(\cos\theta_{23})$,
the leading terms turn to
\begin{align}
	J_{CP}^l
	&\simeq
	-\frac{1}{8\Delta_e}
	a_Eb_E
	d_E^2(c_E'^2 + d_E^2)\sin(2\theta_{12})\left[\cos\theta_{23}-\cos(3\theta_{23})\right]\nonumber\\
	&=
	-\frac{1}{2\Delta_e}
	a_Eb_E
	d_E^2(c_E'^2 + d_E^2)
	\sin(2\theta_{12})\cos\theta_{23}\sin^2\theta_{23}\sin\phi~.
\end{align}
As a typical benchmark, we put the tri-bimaximal mixing \cite{Harrison:2002er,Harrison:2002kp}: $\sin\theta_{12}=1/\sqrt{3}$ and $\sin\theta_{23}=1/\sqrt{2}$.
Then, we have
\begin{align}
	J_{CP}^l
	\simeq
	-\frac{1}{6\Delta_e} a_Eb_E(c_E'^2 + d_E^2)d_E^2\sin\phi
	= -\frac{1}{6\Delta_e} a_Db_Dd_D^2(C_{c'}^2c_D'^2 + C_d^2d_D^2)C_aC_b C_d^2\sin\phi ~.
\end{align}
We can predict $J_{CP}^l$ by inputting the numerical values in Eq.(\ref{bench}) for the parameters $a_D$, $b_D$, $c'_D$, $d_D$, $\phi$; and the CG coefficients from Eq.(\ref{CG-Pati}).
If we choose CG coefficients as follows:
\begin{eqnarray}
	C_a=2 , \quad  C_{a'}=1,
	\quad  C_b=-3,  \quad  C_{c}=\frac{3}{4}, \quad  C_{c'}=1, \quad  C_d=1 ~,
\end{eqnarray}
we obtain
\begin{align}
	&J_{CP}^l
	\simeq
	-0.76 \times 10^{-2} ~ ,\qquad\qquad\
	\frac{m_e^2 ~m_\mu^2~ m_\tau^2 }{m_d^2~ m_s^2 ~m_b^2 }=4~
	({\rm obs:}1.7-7.3) ,
	\nonumber \\
	& \frac{m_e^2 m_\mu^2+ m_\mu^2 m_\tau^2 +m_e^2 m_\tau^2 }
	{m_d^2 m_s^2+ m_s^2 m_b^2 +m_d^2 m_b^2} =24 ~({\rm obs:}15-26),
	~~
	\frac{m_e^2+ m_\mu^2 +m_\tau^2 }
	{m_d^2 + m_s^2  + m_b^2} =1.0 ~({\rm obs:}0.99-1.1)~ ,
\end{align}
which are almost consistent with the observed values.
We must choose the CG coefficients from Eq.\eqref{CG-Pati} so as to reproduce the minus sign of the expected $J_{CP}^l$ in Eq.(\ref{JCPl}).

We can also obtain the mixing angles of the PMNS matrix:
\begin{align}
	\sin^2 \theta_{12}^{\rm PMNS}\simeq 0.38 , \qquad  \sin^2\theta_{23}^{\rm PMNS} \simeq 0.47, \qquad \sin \theta_{13}^{\rm PMNS}\simeq 0.06,\qquad
	\delta_{CP}^l\simeq  -30^\circ .
\end{align}
The predicted value of $\sin^2 \theta_{23}^{\rm PMNS}$ is consistent with the observed value $[0.381,0.615]$($3\sigma$) \cite{NuFIT}.
However, the predicted $\sin^2 \theta_{12}^{\rm PMNS}$ is  little bit larger than the observed value $[0.250,0.354]$($3\sigma$) \cite{NuFIT},
while $\sin \theta_{13}^{\rm PMNS}$ is about a half of observed value $[0.138,0.155]$($3\sigma$) \cite{NuFIT}.
Thus, it is understandable that the predicted $J_{CP}^l$ is $-0.76\times 10^{-2}$ and it is smaller than  the expected value:
 $-2\times 10^{-2}$.

It should be commented that  above data of  mixing angles are presented at the EW scale, but our predictions are given  at the GUT scale.
The more accurate study including the RG evolution of the mixing angles is necessary for further discussions.

We present the summary of $\mathrm{Tr}([H_\nu,H_E]^3)$ in the case where the neutrino mass matrix leads to the tri-bimaximal mixing in Table 3.
It is remarked that all $J_{CP}^l$ have the same sign as the quark CP violation $J_{CP}^q$
as far as the sign of CG coefficients are positive.
Therefore, one negative CG coefficient $-3$ should be taken at least.
The magnitudes of $J_{CP}^l$ for all cases of textures can be consistent with the expected value  in order-of-magnitude estimate while the proper mass ratios between down-type quarks and charged leptons are reproduced.


\subsubsection{SU(5) model with non-diagonal neutrino mass matrix}

We discuss the case of the SU(5) model with the charged lepton mass matrix $M_E^{(1)}$ in Eq.(\ref{charged-1-SU5}) and the neutrino mass matrix $M_\nu$ in Eq.(\ref{neutrinomassmatrix}).
By using Eq.(\ref{eq:Tr_nu_E}), we have
\begin{align}
	\mathrm{Tr}([H_\nu , H_E^{(1)}]^3)
	&=-6iJ_{CP}^l ~\Delta_{\nu} ~\Delta_{e}
	\nonumber\\
	&=
	-\frac{3}{4}i \Delta_{\nu}  a_E'b_E
	\left[
	4a_E'c_E\cos(2\theta_{12})
	\left\{
	2c_E'd_E\cos(2\theta_{23})
	-
	(a_E^2 + c_E'^2 -d_E^2)\sin(2\theta_{23})
	\right\}\right.\nonumber\\
	&\phantom{=}\left.
	+\sin(2\theta_{12})
	\left\{
	-\left[
	(a_E'^2-4c_E^2)(a_E^2+c_E'^2)
	-
	(a_E^2 + a_E'^2 +b_E^2 -c_E^2 -c_E'^2)d_E^2 + d_E^4
	\right]\cos\theta_{23}\right.\right.\nonumber\\
	&\phantom{=}\left.\left.
	+
	\left[
	a_E'^2(a_E^2 + c_E'^2)
	-
	(a_E^2 + a_E'^2 + b_E^2 -c_E^2 + 3c_E'^2)d_E^2 + d_E^4
	\right]\cos(3\theta_{23})\right.\right.\nonumber\\
	&\phantom{=}\left.\left.
	+c_E'd_E(a_E^2 - 2a_E'^2 + b_E^2 +3c_E^2 +c_E'^2 +d_E^2)\sin\theta_{23}
	\right.\right.\nonumber\\
	&\phantom{=}\left.\left.
	-8b_Ec_Ed_E(c_E'\cos\theta_{23} + d_E\sin\theta_{23})\cos^2\theta_{23}\cos\phi
	\right.\right.\nonumber\\
	&\phantom{=}\left.\left.
	+
	c_E'd_E(a_E^2 + 2a_E'^2+b_E^2 -c_E^2 +c_E'^2 -3d_E^2)\sin(3\theta_{23})
	\right\}
	\right]\sin\phi~.
\end{align}
We express $J_{CP}^l$ in the leading order  as:
\begin{align}
	J_{CP}^l
	&\simeq
	\frac{1}{8\Delta_{e}}
	a_E'b_E
	\left[
	4a_E'c_E\cos(2\theta_{12})
	\left\{
	2c_E'd_E\cos(2\theta_{23})
	-
	(c_E'^2 -d_E^2)\sin(2\theta_{23})
	\right\}\right.\nonumber\\
	&\phantom{=}\left.
	+\sin(2\theta_{12})
	\left\{
	-(c_E'^2+d_E^2)d_E^2\cos\theta_{23}
	-(3c_E'^2 - d_E^2)d_E^2\cos(3\theta_{23})
	\right.\right.\nonumber\\
	&\phantom{=}\left.\left.
	+c_E'd_E(c_E'^2 +d_E^2)\sin\theta_{23}
	-8b_Ec_Ed_E(c_E'\cos\theta_{23} + d_E\sin\theta_{23})\cos^2\theta_{23}\cos\phi
	\right.\right.\nonumber\\
	&\phantom{=}\left.\left.
	+
	c_E'd_E(c_E'^2 -3d_E^2)\sin(3\theta_{23})
	\right\}
	\right]\sin\phi~.
\end{align}
If we suppose $\mathcal{O}(\sin\theta_{12})\sim\mathcal{O}(\cos\theta_{12})$ and
$\mathcal{O}(\sin\theta_{23})\sim\mathcal{O}(\cos\theta_{23})$,
$J_{CP}^l$ is given as:
\begin{align}
	J_{CP}^l
	&\simeq
	\frac{1}{8\Delta_e}
	a_E'b_E
	\sin(2\theta_{12})
	\left\{
	-(c_E'^2+d_E^2)d_E^2\cos\theta_{23}
	-(3c_E'^2 - d_E^2)d_E^2\cos(3\theta_{23})
	\right.\nonumber\\
	&\phantom{=}\left.
	+c_E'd_E(c_E'^2 +d_E^2)\sin\theta_{23}
	+
	c_E'd_E(c_E'^2 -3d_E^2)\sin(3\theta_{23})
	\right\}\sin\phi~ \nonumber\\
	&=
	\frac{1}{2\Delta_e}~
	a_E'b_E
	d_E \sin(2\theta_{12})(c_E'\sin\theta_{23}-d_E\cos\theta_{23})
	(c_E'\cos\theta_{23} + d_E\sin\theta_{23})^2\sin\phi~ .
\end{align}
In the tri-bimaximal mixing basis,
$\sin\theta_{12}=1/\sqrt{3}$ and $\sin\theta_{23}=1/\sqrt{2}$, the approximated Jarlskog invariant $J_{CP}^l$ becomes
\begin{align}
	J_{CP}^l &\simeq
	-\frac{1}{6\Delta_e}
	a_E'b_E d_E
	(c_E' + d_E)
	(d_E^2-c_E'^2)
	\sin\phi~  \nonumber \\
	&\simeq \frac{1}{6 m_\mu^2 m_\tau^4}
	a_D'b_D d_D
	(C_{c'} c_D' + C_d d_D)
	(C_d^2 d_D^2- C_{c'}^2c_D'^2) C_{a'} C_b C_d \sin\phi ~.
\end{align}
We can predict $J_{CP}^l$ by inputting the numerical values in Eq.(\ref{bench}) for the parameters $a'_D$, $b_D$, $c'_D$, $d_D$, and $\phi$; and the CG coefficients from Eq.(\ref{CG-SU5}).
We use the following choice of CG coefficients as:
\begin{eqnarray}
	C_a=1 , \quad  C_{a'}=\frac{9}{2},
	\quad  C_b=\pm\frac{9}{2}, \quad C_c=\frac{9}{2},  \quad   C_{c'}=-\frac{3}{2}, \quad  C_d=-\frac{1}{2} ~ ~.
\end{eqnarray}
Then, we obtain
\begin{align}
	&J_{CP}^l
	\simeq
	-1.13\times 10^{-2} ~ ,\qquad\qquad
	\frac{m_e^2 ~m_\mu^2~ m_\tau^2 }{m_d^2~ m_s^2 ~m_b^2 }=5.06~
	({\rm obs:}1.7-7.3) ,
	\nonumber \\
	& \frac{m_e^2 m_\mu^2+ m_\mu^2 m_\tau^2 +m_e^2 m_\tau^2 }
	 {m_d^2 m_s^2+ m_s^2 m_b^2 +m_d^2 m_b^2} =26 ~({\rm obs:}15-26),
	 \qquad
	 \frac{m_e^2+ m_\mu^2 +m_\tau^2 }
	 {m_d^2 + m_s^2  + m_b^2} =1.07~({\rm obs:}0.99-1.1) ~ .
\end{align}
Thus, the expected value of $J_{CP}^l$ in Eq.(\ref{JCPl}) is easily reproduced up to its sign by taking the relevant CG coefficients.
We also show the mixing angles of the PMNS matrix numerically:
\begin{align}
	\sin^2 \theta_{12}^{\rm PMNS}\simeq 0.28 , \qquad  \sin^2\theta_{23}^{\rm PMNS} \simeq 0.85, \qquad
	\sin \theta_{13}^{\rm PMNS}\simeq 0.153 ,\qquad \delta_{CP}^l\simeq  -113^\circ .
\end{align}
The predicted value of $\sin^2 \theta_{12}^{\rm PMNS}$ is consistent with the observed value $[0.250,0.354]$($3\sigma$) \cite{NuFIT}.
The predicted  $\sin \theta_{13}^{\rm PMNS}$ is also consistent with  the observed value $[0.138,0.155]$($3\sigma$) \cite{NuFIT}.
However, $\sin^2 \theta_{23}^{\rm PMNS}$ is rather larger than the observed value $[0.381,0.615]$($3\sigma$) \cite{NuFIT}.
Since  above data of  mixing angles are presented at the EW scale, the study including the RG evolution of the mixing angles is necessary in further discussions.

We have presented the summary of $\mathrm{Tr}([H_\nu,H_E]^3)$ in the case where the neutrino mass matrix leads to the tri-bimaximal mixing in Table 3.
It is noticed that the sign of $J_{CP}^l$ depends on the textures apart from the sign of CG coefficients.
Since there are several kinds of the CG coefficients with minus sign, it is easy to predict the successful  $J_{CP}^l$  in order-of-magnitude estimate
with the favorable mass relations between the down-type quarks and charged leptons.

\begin{table}[htb]
	\begin{center}
		\begin{tabular}{ccc}
			\hline
			$M_D$
			& $\tr\left([H_\nu,H_E]^3\right)_{\rm Pati-Salam}$
			&$\tr\left([H_\nu,H_E]^3\right)_{\rm SU(5)}$
			\\[2pt]
			\hline
			$M_D^{(1)}$
			& $ia_Eb_E(c_E'^2+d_E^2)d_E^2\Delta_\nu\sin\phi$
			&  $ia_E'b_E(c_E'+d_E)d_E(d_E^2-c_E'^2)\Delta_\nu\sin\phi$
			\\[4pt]
			$M_D^{(2)}$
			& $ia_Eb_E(c_E'^2+d_E^2)d_E^2\Delta_\nu\sin\phi$
			& $ia_Ea_E'b_Ec_E(c_E'+d_E)^2\Delta_\nu\sin\phi$
			\\[4pt]
			$M_D^{(3)}$
			& $ia_Eb_E(c_E'^2+d_E^2)d_E^2\Delta_\nu\sin\phi$
			&  $-ia_E'b_Ec_E(d_E^2-c_E'^2)(c_E+d_E)\Delta_\nu\sin\phi$
			\\[4pt]
			$M_D^{(4)}$
			& $ia_Eb_Ed_E^4\Delta_\nu\sin\phi$
			&  $ia_E'b_Ed_E^4\Delta_\nu\sin\phi$
			\\[4pt]
			$M_D^{(5)}$
			& $ia_Eb_Ed_E^4\Delta_\nu\sin\phi$
			& $ia_E'b_Ec_E(a_E-c_E')d_E^2\Delta_\nu\sin\phi$
			\\[4pt]
			$M_D^{(6)}$
			& $ia_Eb_Ed_E^4\Delta_\nu\sin\phi$
			& $-ia_E'b_Ec_Ed_E^3\Delta_\nu\sin\phi$
			\\[4pt]
			$M_D^{(11)}$
			& $ia_Ec_Ec_E'd_E(c_E'^2+d_E^2)\Delta_\nu\sin\phi$
			& $ia_Ea_E'd_E(c_E'd_E^2+d_E^3-c_E'^3-c_E'^2d_E)\Delta_\nu\sin\phi$
			\\[4pt]
			$M_D^{(12)}$
			& $ia_Ec_Ec_E'd_E(c_E'^2+d_E^2)\Delta_\nu\sin\phi$
			& $-ia_Ea_E'b_Ec_E(c_E'+d_E)^2\Delta_\nu\sin\phi$
			\\[4pt]
			$M_D^{(13)}$
			& $ia_Ec_Ec_E'd_E(c_E'^2+d_E^2)\Delta_\nu\sin\phi$
			&  $-ia_Ea_E'b_E(d_E^2-c_E'^2)(c_E'+d_E)\Delta_\nu\sin\phi$
			\\[4pt]
			$M_D^{(14)}$
			& $ia_Eb_Ed_E^4\Delta_\nu\sin\phi$
			& $ia_Ea_E'd_E^4\Delta_\nu\sin\phi$
			\\[4pt]
			$M_D^{(15)}$
			& $ia_Ec_Ec_E'd_E(c_E'^2+d_E^2)\Delta_\nu\sin\phi$
			& $ia_Ea_E'd_E^2(c_E'd_E+2c_E'^2-d_E^2)\Delta_\nu\sin\phi$
			\\[4pt]
			$M_D^{(16)}$
			& $ib_Ec_Ec_E'^2(c_E'^2+d_E^2)\Delta_\nu\sin\phi$
			& $ia_E'c_E\left |a_E'd_E-c_Ec_E'e^{i\phi }
			\right |^2\Delta_\nu\sin\phi$
			\\[4pt]
			$M_D^{(17)}$
			& $ia_Ec_Ec_E'd_E(c_E'^2+d_E^2)\Delta_\nu\sin\phi$
			& $-ib_Ec_E\left |c_Ec_E'-b_Ed_Ee^{i\phi }
			\right |^2\Delta_\nu\sin\phi$
			\\[4pt]
			\hline
		\end{tabular}
		\caption{Summary of $\tr\left([H_\nu,H_E]^3\right)$ for the Pati--Salam model and the SU(5) model in the case where the neutrino mass matrix leads to the tri-bimaximal mixing.
			Note that $\tr\left([H_\nu,H_E]^3\right)=
			-6i\Delta_\nu \Delta_e J_{\mathrm{CP}}^l$.}
	\end{center}
\end{table}

\section{Summary}

We have investigated the relation between CP violations in the quark sector and that in the lepton sector,
especially for the sign of the CP violating Dirac phase in the Pati--Salam model and the SU(5) model.
In the standpoint of  ``Occam's Razor'',
we have considered the down-type quark mass matrix with three zeros while the up-type quark mass matrix is diagonal,
which leads to the successful CKM mixing angles and CP violation.
The down-type quark mass matrix has six real parameters and one phase which is the only source of the CP violation in the quark sector.
The thirteen textures of down-type quark mass matrix are consistent with the present experimental data of quark masses and CKM parameters.

These successful quark mass matrices  are embedded in
the Pati--Salam model and the SU(5) model where one single operator dominates each matrix element.
Then, the down-type quark mass matrix is related to the charged lepton mass matrix.
The charged lepton mass matrix is obtained from the down-type quark mass matrix with CG coefficients of the symmetry which are necessary to reproduce the observed mass ratios of quarks and leptons.
Therefore, the CP violating phase in the down-type quark mass matrix also appears in the charged lepton mass matrix.
Our investigation is taken place for both two cases:
the neutrino mass matrix is diagonal or non-diagonal,
where no additional CP violating phase is introduced apart from the Majorana phases.

In the case of the diagonal neutrino mass matrix,
we have presented a lepton mixing (PMNS) matrix for a typical texture of quarks and leptons for both Pati--Salam and SU(5) models.
The minus sign of $J_{CP}^l$ ($\delta_{CP}^l$) and the proper mass ratios between down-type quarks and charged leptons
are obtained by choosing relevant CG coefficients.
However, the predicted magnitude of Jarlskog invariant is
at most ${\cal O}(10^{-4})$ which is two order smaller than the expected value of the observation $-0.02$.
The other twelve textures of the quark and lepton mass matrices are in the same situations as the above texture.

We have also discussed $J_{CP}^l$ in the case of the non-diagonal neutrino mass matrix
where a new CP violating phase does not appear apart from the Majorana phases.
We have taken the tri-bimaximal mixing pattern of the neutrino mass matrix as a benchmark and investigated the magnitude of $J_{CP}^l$.
For the Pati--Salam model, a negative CG coefficient $-3$ is required to reproduce the minus sign of $J_{CP}^l$ for the all cases of textures.
We have predicted the proper mass ratios of down-type quarks and charged leptons as well as the magnitudes of $J_{CP}^l$ which is almost consistent with the observed one.
For the SU(5) model, we have shown the successful prediction of $J_{CP}^l$
by choosing the relevant CG coefficients in a typical texture of the mass matrix.
The sign of $J_{CP}^l$ is different for the textures in the SU(5) model as seen in Table 3.
The precise measurement of $J_{CP}^l$ including its sign provides an important key towards the quark-lepton unification in GUT.

Our predictions have been given at the GUT scale.
The systematic study including the RG evolution from the EW scale to the  GUT scale will appear elsewhere.


\vspace{1 cm}
\noindent
{\bf Acknowledgement}

This work is  supported by JSPS Grand-in-Aid for Scientific Research
16J05332 (YS) and  15K05045 (MT).

\vskip 1 cm
\appendix

\section*{Appendix}

\section{Redundancy of our textures}

Since the CKM matrix is the flavor mixing among  the left-handed quarks,
the textures in Eqs.(\ref{downmassmatrixA}) and (\ref{downmassmatrixB}) have freedoms of the unitary transformation of the right-handed quarks.
Under such a transformation, $M_D M_D^\dagger$ is invariant.
We can easily find that some textures are transformed into other ones as follows:

\begin{eqnarray}
	M_D^{(2)} \equiv M_D^{(16)} \equiv M_D^{(17)}\ ,  \qquad
	M_D^{(5)}\equiv M_D^{(14)}  \ ,  \qquad
	M_D^{(11)} \equiv M_D^{(15)}   \ ,
\end{eqnarray}
where the notation $"\equiv"$ means the equivalence up to the unitary transformation of the right-handed quarks.

\section{Masses and  CKM parameters for  $M_D^{(1)}$}

We show how to predict the CKM mixing angles and the CP violation for the case of $M_D^{(1)}$ in  Eq.(\ref{downmassmatrixA}) as a representative.
Since the up-type quark mass matrix is diagonal, the CKM matrix is obtained as a diagonalizing matrix of the down-type quark mass matrix $M_D^{(1)}$.
In order to determine the left-handed quark mixing angles, we diagonalize $M_D^{(1)} M_D^{(1)\dagger}$;
\begin{equation}
	M_D^{(1)} M_D^{(1)\dagger}=
	\begin{pmatrix}
	a^2_D & a_D b_D\  e^{i\phi} & a_D c'_D \\
	a_D b_D\  e^{-i\phi} &a'^2_D+b_D^2+c_D^2& b_D c'_D\  e^{-i\phi}+c_D d_D\\
	a_D c'_D &  b_D c'_D\  e^{i\phi}+c_D d_D & c'^2_D+d^2_D
	\end{pmatrix}_{LL} \ .
\label{MMdagger}
\end{equation}
By solving the eigenvalue equation of  $M_D^{(1)} M_D^{(1)\dagger}$,
we obtain the following relations between the down-type quark masses and the parameters:
\begin{align}
	&m_d^2+m_s^2+m_b^2=a^2_D+a'^2_D+b^2_D+c^2_D+c'^2_D+d^2_D \ , \nonumber \\
	&m_d^2 m_s^2+ m_s^ 2m_b^2+m_b^2 m_d^2=
	a^2_D a'^2_D +a^2_D (c^2_D+ d^2_D)+a'^2_D (c'^2_D+d^2_D)
	+c^2_D c'^2_D+b^2_D d^2_D -2b_Dc_D c'_D d_D\cos\phi\ ,   \nonumber \\
	&m_d^2 m_s^2 m_b^2= a^2_D a'^2_D d^2_D \ .
\label{massrelations}
\end{align}
Moreover, the eigenvectors lead to the CKM matrix elements $V_{ij}$ and the CKM phase $\delta_{CP}^q$ which are given in accordance with the PDG parametrization \cite{Tanabashi:2018oca}.
Those of the leading order are
\begin{equation}
	|V_{us}|\simeq \frac{a_D b_D}{m_s^2}\left | \sin\frac{\phi}{2} \right | \ ,
	\quad
	|V_{cb}|\simeq \sqrt{2}\frac{c_D}{m_b} \left |\cos\frac{\phi}{2} \right |\ ,
	\quad
	|V_{ub}|\simeq \frac{a_D c'_D}{m_b^2} \ ,
	\quad
	\delta_{CP}^q\simeq \frac{1}{2}(\pi-\phi) \ ,
\label{CKM}
\end{equation}
where we adopt the approximate relations $b_D\sim c_D$ and $c'_D\sim d_D$, which will be justified in our numerical results.
It is noted that the next-to-leading order correction is included for $\delta_{CP}^q$.

The CP violating measure, the Jarlskog invariant $J_{CP}^q$ \cite{Jarlskog:1985ht}, is derived from the following relation \cite{Bernabeu:1986fc,Gronau:1986xb,Branco:1999fs,Branco:2011zb}:
\begin{align}
	{\rm Tr}([H_U, H_D]^3) = 6 i J_{CP}^q   \Delta_u \Delta_d \ ,
	\label{Jcp}
\end{align}
where
\begin{align}
	H_U=M_U M_U^\dagger  \ , \qquad\qquad H_D=M_D M_D^\dagger ~ .
\end{align}
Instead of $J^q_{CP}$, we can use another form of $J_{CP}^q$:
\begin{align}
	j^q_{CP}\equiv  - \Delta_d  J^q_{CP}  \ .
\end{align}
In the case of $M_D^{(1)}$, we can express $J_{CP}^q$ in terms of the mass matrix elements:
\begin{align}
	j^q_{CP}={a^2_D b_D c_D c'_D d_D \sin\phi} \ .
\end{align}

\begin{table}[hbp]
	\begin{center}
		\begin{tabular}{c|c|c|c|c|c|c|c}
			\hline
			& $a_D$ [MeV]& $a'_D$ [MeV] & $b_D$ [MeV] &  $c_D$ [MeV]& $c'_D$ [GeV]& $d_D$ [GeV] &$\phi$ [$\circ$]\\
			\hline\hline
			$M_{D}^{(1)}$ &$15$-$17.5$&$10$-$15$&$92$-$104$& $78$-$95$&$1.65$-$2.0$ & $2.0$-$2.3$ &$37$-$48$\\
			\hline
			$M_{D}^{(2)}$&$15$-$17$ &$2$-$4$ & $94$-$106$&$78$-$95$&$1.65$-$2.0$ &$2.0$-$2.3$ &$40$-$49$\\
			\hline
			$M_{D}^{(3)}$&$15$-$17.5$ &$250$-$380$&$92$-$104$&$78$-$95$&$1.65$-$2.0$ &$2.0$-$2.3$ &$37$-$48$\\
			\hline
			$M_{D}^{(4)}$&$11$-$14$ &$9$-$17$&$45$-$58$&$115$-$128$&$0.009$-$0.011$ &$2.8$-$2.9$ &$63$-$75$\\
			\hline
			$M_{D}^{(5)}$&$11$-$14$ &$2$-$4$&$45$-$58$&$115$-$128$&$0.009$-$0.011$ &$2.8$-$2.9$ &$63$-$75$\\
			\hline
			$M_{D}^{(6)}$&$11$-$14$ &$220$-$420$&$45$-$58$&$115$-$128$&$0.009$-$0.011$ &$2.8$-$2.9$ &$63$-$75$\\
			\hline
			$M_{D}^{(11)}$ &$10$-$12$&$2.5$-$3.5$&$11$-$13$&$125$-$135$&$1.0$-$1.2$ & $2.5$-$2.7$ &$104$-$118$\\
			\hline
			$M_{D}^{(12)}$ &$10$-$12$&$11$-$18$&$11$-$13$& $125$-$135$&$1.0$-$1.2$ & $2.5$-$2.7$ &$106$-$120$\\
			\hline
			$M_{D}^{(13)}$ &$10$-$12$&$260$-$390$&$11$-$13$& $125$-$135$&$1.0$-$1.2$ & $2.5$-$2.7$ &$104$-$118$\\
			\hline
			$M_{D}^{(14)}$&$11$-$14$ &$2$-$4$&$45$-$58$&$115$-$128$&$0.009$-$0.011$&$2.8$-$2.9$ &$63$-$75$\\
			\hline
			$M_{D}^{(15)}$&$10$-$12$&$2.5$-$3.5$ &$11$-$13$&$125$-$135$&$1.0$-$1.2$&$2.5$-$2.7$ &$104$-$118$\\
			\hline
			$M_{D}^{(16)}$&$2$-$4$&$78$-$95$ &$15$-$17$&$94$-$106$&$2.0$-$2.3$&$1.65$-$2.0$ &$40$-$49$\\
			\hline
			$M_{D}^{(17)}$&$15$-$17$ &$2$-$4$&$94$-$106$&$78$-$95$&$1.65$-$2.0$&$2.0$-$2.3$ &$40$-$49$\\
			\hline
		\end{tabular}
		\caption{The allowed regions of the parameters for each $M_D^{(k)}$ $(k=1\,\mbox{--}\,6, 11\,\mbox{--}\,17)$
			at the EW scale\cite{Tanimoto:2016rqy}.}
		\label{tab4}
	\end{center}
\end{table}

It is convenient to eliminate the parameters $a'_D$, $d_D$, and $\phi$ by using Eq.(\ref{massrelations}) in the numerical calculations.
As an input, we adopt the data of the down-type quark Yukawa couplings at the EW scale $M_Z$ by taking $90\%$ C.L \cite{Antusch:2013jca} as
\begin{eqnarray}
	y_d=(1.58^{+0.23}_{-0.10}) \times 10^{-5}, \quad
	y_s=(3.12^{+0.17}_{-0.16}) \times 10^{-4}, \quad
	y_b=(1.639\pm 0.015) \times 10^{-2},
\label{dyukawa}
\end{eqnarray}
which give down-type quark masses as $m_q=y_q v_H ~(q=d,s,b)$ with $v_H=174.1$ GeV.

After inputting the Yukawa couplings of down-type quarks, we have four parameters $a_D$, $b_D$,  $c_D$, and $c'_D$, which are determined by the four CKM parameters.
The allowed region of parameters are listed  in Table \ref{tab4} \cite{Tanimoto:2016rqy}.

In order to perform numerical discussions at the GUT scale,
we adopt the following inputs, down-type quark Yukawa couplings and CKM parameters, at the GUT scale $2\times 10^{16}$GeV with $\tan\beta=10$ in the framework of MSSM \cite{Bjorkeroth:2015ora}:
\begin{align}
&y_d=(4.84\pm 1.07) \times 10^{-6}, \quad y_s=(9.59\pm 1.04) \times 10^{-5}, \quad y_b=(7.01\pm 0.178) \times 10^{-3}, \nonumber \\
&\theta_{12}^{\mathrm{CKM}}=(13.027\pm 0.0814)^\circ,
\quad~~ \theta_{23}^{\mathrm{CKM}}=(2.054\pm 0.384)^\circ,
 \quad\quad~ \theta_{13}^{\mathrm{CKM}}=(0.1802\pm 0.0281)^\circ, \nonumber\\
& \delta_{CP}^q=(69.21\pm 6.19)^\circ~,
\end{align}
where PDG notations are used \cite{Tanabashi:2018oca}.
We  also show the charged lepton Yukawa couplings at the GUT scale
 with $\tan\beta=10$ \cite{Bjorkeroth:2015ora}:
\begin{eqnarray}
y_e=(1.98\pm 0.024) \times 10^{-6}, \qquad
y_\mu=(4.19\pm 0.050) \times 10^{-4}, \qquad
y_\tau=(7.15\pm 0.074) \times 10^{-3},
\end{eqnarray}
where the charged lepton masses are given  as $m_l=y_l  v_H$.


\section{Charged lepton mass matrices}
We summarize the charged lepton mass matrix in both the Pati--Salam and SU(5) models.
\subsection{Pati--Salam model}
\begin{align}
	\begin{aligned}
	M_E^{(1)} =&
	\begin{pmatrix}
	0 & a_E & 0 \\
	a'_E & b_E\,e^{-i\phi} & c_E \\
	0 & c'_E & d_E
	\end{pmatrix}_{LR}, ~
	M_E^{(2)} =
	\begin{pmatrix}
	a'_E & a_E & 0 \\
	0 & b_E\,e^{-i\phi} & c_E \\
	0 & c'_E & d_E
	\end{pmatrix}_{LR},~
	M_E^{(3)} =
	\begin{pmatrix}
	0 & a_E & 0 \\
	0 & b_E\,e^{-i\phi} & c_E \\
	a'_E & c'_E & d_E
	\end{pmatrix}_{LR}, \\
	M_E^{(4)} =&
	\begin{pmatrix}
	0 & a_E & c'_E \\
	a'_E & b_E\,e^{-i\phi} & c_E \\
	0 & 0 & d_E
	\end{pmatrix}_{LR}, ~
	M_E^{(5)} =
	\begin{pmatrix}
	a'_E & a_E & c'_E \\
	0 & b_E\,e^{-i\phi} & c_E \\
	0 & 0 & d_E
	\end{pmatrix}_{LR},~
	M_E^{(6)} =
	\begin{pmatrix}
	0 & a_E & c'_E \\
	0 & b_E\,e^{-i\phi} & c_E \\
	a'_E & 0 & d_E
	\end{pmatrix}_{LR},\\
	M_E^{(11)} =&
	\begin{pmatrix}
	a'_E & a_E\,e^{-i\phi} & b_E \\
	0 & 0 & c_E \\
	0 & c'_E & d_E
	\end{pmatrix}_{LR},~~
	M_E^{(12)} =
	\begin{pmatrix}
	0 & a_E\,e^{-i\phi} & b_E \\
	a'_E & 0 & c_E \\
	0 & c'_E & d_E
	\end{pmatrix}_{LR},~~
	M_E^{(13)} =
	\begin{pmatrix}
	0 & a_E\,e^{-i\phi} & b_E \\
	0 & 0 & c_E \\
	a'_E & c'_E & d_E
	\end{pmatrix}_{LR}, \\
	M_E^{(14)} =&
	\begin{pmatrix}
	a_E\,e^{i\phi} & a'_E & c'_E \\
	b_E & 0 & c_E \\
	0 & 0 & d_E
	\end{pmatrix}_{LR},~~
	M_E^{(15)} =
	\begin{pmatrix}
	a_E\,e^{-i\phi} & a'_E & b_E \\
	0 & 0 & c_E \\
	c'_E & 0 & d_E
	\end{pmatrix}_{LR}, \\
	M_E^{(16)} =&
	\begin{pmatrix}
	0 & a_E & b_E \\
	a'_E & 0 & c_E\,e^{-i\phi} \\
	c'_E & 0 & d_E
	\end{pmatrix}_{LR},~~
	M_E^{(17)} =
	\begin{pmatrix}
	a_E & a'_E & 0 \\
	b_E & 0 & c_E\,e^{i\phi} \\
	c'_E & 0 & d_E
	\end{pmatrix}_{LR}.
	\end{aligned}
\end{align}
\subsection{SU(5) model}
\begin{align}
	\begin{aligned}
	M_E^{(1)} =&
	\begin{pmatrix}
	0 & a'_E & 0 \\
	a_E & b_E\,e^{-i\phi} & c'_E \\
	0 & c_E & d_E
	\end{pmatrix}_{LR}, ~~
	M_E^{(2)} =
	\begin{pmatrix}
	a'_E & 0 & 0 \\
	a_E & b_E\,e^{-i\phi} & c'_E \\
	0 & c_E & d_E
	\end{pmatrix}_{LR}, ~~
	M_E^{(3)} =
	\begin{pmatrix}
	0 & 0 & a'_E \\
	a_E & b_E\,e^{-i\phi} & c'_E \\
	0 & c_E & d_E
	\end{pmatrix}_{LR}, \\
	M_E^{(4)} =&
	\begin{pmatrix}
	0 & a'_E & 0 \\
	a_E & b_E\,e^{-i\phi} & 0 \\
	c'_E & c_E & d_E
	\end{pmatrix}_{LR}, ~~
	M_E^{(5)} =
	\begin{pmatrix}
	a'_E & 0 & 0 \\
	a_E & b_E\,e^{-i\phi} & 0 \\
	c'_E & c_E & d_E
	\end{pmatrix}_{LR}, ~~
	M_E^{(6)} =
	\begin{pmatrix}
	0 & 0 & a'_E \\
	a_E & b_E\,e^{-i\phi} & 0 \\
	c'_E & c_E & d_E
	\end{pmatrix}_{LR},\\
	M_E^{(11)} =&
	\begin{pmatrix}
	a'_E & 0 & 0 \\
	a_E\,e^{-i\phi} & 0 & c'_E \\
	b_E & c_E & d_E
	\end{pmatrix}_{LR}, ~~
	M_E^{(12)} =
	\begin{pmatrix}
	0 & a'_E & 0 \\
	a_E\,e^{-i\phi} & 0 & c'_E \\
	b_E & c_E & d_E
	\end{pmatrix}_{LR}, ~~
	M_E^{(13)} =
	\begin{pmatrix}
	0 & 0 & a'_E \\
	a_E\,e^{-i\phi} & 0 & c'_E \\
	b_E & c_E & d_E
	\end{pmatrix}_{LR}, \\
	M_E^{(14)} =&
	\begin{pmatrix}
	a_E\,e^{i\phi} & b_E & 0 \\
	a'_E & 0 & 0 \\
	c'_E & c_E & d_E
	\end{pmatrix}_{LR}, ~~~
	M_E^{(15)} =
	\begin{pmatrix}
	a_E\,e^{-i\phi} & 0 & c'_E \\
	a'_E & 0 & 0 \\
	b_E & c_E & d_E
	\end{pmatrix}_{LR}, \\
	M_E^{(16)} =&
	\begin{pmatrix}
	0 & a'_E & c'_E \\
	a_E & 0 & 0 \\
	b_E & c_E\,e^{-i\phi} & d_E
	\end{pmatrix}_{LR}, ~~
	M_E^{(17)} =
	\begin{pmatrix}
	a_E & b_E & c'_E \\
	a'_E & 0 & 0 \\
	0 & c_E\,e^{i\phi} & d_E
	\end{pmatrix}_{LR}.
	\end{aligned}
\end{align}

\section{Formulae of CP violation}
We present a brief review of the formulae of the CP violation \cite{Bernabeu:1986fc,Gronau:1986xb,Branco:1999fs,Branco:2011zb}.
The CP transformations with $n_g$ generations are given as:
\begin{align}
	(\mathcal{CP})u_L(t,\boldsymbol{r})(\mathcal{CP})^{-1}
	&=
	K_L^q\gamma^0 C \overline{u_L}^T(t,-\boldsymbol{r})~,\label{eq:CP_pL}\\
	(\mathcal{CP})d_L(t,\boldsymbol{r})(\mathcal{CP})^{-1}
	&=
	K_L^q\gamma^0 C \overline{d_L}^T(t,-\boldsymbol{r})~,\\
	(\mathcal{CP})u_R(t,\boldsymbol{r})(\mathcal{CP})^{-1}
	&=
	K_R^u\gamma^0 C \overline{u_R}^T(t,-\boldsymbol{r})~,\\
	(\mathcal{CP})d_R(t,\boldsymbol{r})(\mathcal{CP})^{-1}
	&=
	K_R^d\gamma^0 C \overline{d_R}^T(t,-\boldsymbol{r})~,
\end{align}
for the quarks in flavor basis and
\begin{align}
	(\mathcal{CP})\nu_L(t,\boldsymbol{r})(\mathcal{CP})^{-1}
	&=
	K_L^l\gamma^0 C \overline{\nu_L}^T(t,-\boldsymbol{r})~,\\
	(\mathcal{CP})e_L(t,\boldsymbol{r})(\mathcal{CP})^{-1}
	&=
	K_L^l\gamma^0 C \overline{e_L}^T(t,-\boldsymbol{r})~,\\
	(\mathcal{CP})e_R(t,\boldsymbol{r})(\mathcal{CP})^{-1}
	&=
	K_R^e\gamma^0 C \overline{e_R}^T(t,-\boldsymbol{r})~,\label{eq:CP_nR}
\end{align}
for the leptons in flavor basis. Here the matrices $K_{L}^{q,l}$ and $K_{R}^{u,d,e}$ are $n_g\times n_g$ unitary matrices which act in generation space and $C$ denotes the charge conjugation operator.
In order to hold invariance under the CP transformations Eqs.\eqref{eq:CP_pL}-\eqref{eq:CP_nR} for the mass terms,
\begin{align}
	\mathcal{L}_M
	=
	-\overline{u_L}M_U u_R - \overline{d_L}M_D d_R
	-\frac{1}{2}\overline{\nu_L}M_\nu (\nu_L)^c
	-\overline{e_L}M_E e_R + \mathrm{h.c.}~,
\end{align}
the mass matrices should satisfy,
\begin{align}
	K_L^{q\dagger} M_U K_R^u
	=
	M_U^\ast~, \quad
	K_L^{q\dagger} M_D K_R^d
	&=
	M_D^\ast~, \quad
	K_L^{l\dagger} M_E K_R^e
	=
	M_E^\ast~,\label{eq:CP_con_UDE}\\
	K_L^{l\dagger} M_\nu K_L^{l\ast}
	&=
	-M_\nu^\ast~.\label{eq:CP_con_nu}
\end{align}
Eqs.\eqref{eq:CP_con_UDE} and \eqref{eq:CP_con_nu} imply
\begin{align}
	K_L^{q\dagger} H_U K_L^q
	=
	H_U^\ast~, \quad
	K_L^{q\dagger} H_D K_L^q
	&=
	H_D^\ast~, \quad
	K_L^{l\dagger} H_E K_L^l
	=
	H_E^\ast~,\label{eq:CP_con_HUDE}\\
	K_L^{l\dagger} H_\nu K_L^l
	&=
	H_\nu^\ast~,
	\label{eq:CP_con_Hnu}
\end{align}
where $H_i = M_iM_i^\dagger~(i=U,D,\nu,E)$.
By using the conditions Eqs.\eqref{eq:CP_con_HUDE} and \eqref{eq:CP_con_Hnu}, we obtain necessary conditions of CP invariance:
\begin{align}
	\tr\left([H_U,H_D]^3\right) = 0~,\quad
	\tr\left([H_\nu,H_E]^3\right) = 0~.\label{eq:CP_inv_cond}
\end{align}
The CP invariance does not hold if the mass matrices do not satisfy the conditions in Eq.\eqref{eq:CP_inv_cond}.
We obtain the following relations by computing $\tr\left([H_U,H_D]^3\right)$ and $\tr\left([H_\nu,H_E]^3\right)$,
\begin{align}
	\tr\left([H_U,H_D]^3\right)
	&=
	6i\sum_{\alpha,\beta = u,c,t}\sum_{i,j=d,s,b}
	m_\alpha^4m_\beta^2m_i^4m_j^2\Im[V_{\alpha i}V_{\beta j}V_{\beta i}^\ast V_{\alpha j}^\ast]~,\label{eq:Trud3}\\
	\tr\left([H_\nu,H_E]^3\right)
	&=
	-6i\sum_{\alpha,\beta = 1,2,3}\sum_{i,j=e,\mu,\tau}
	m_\alpha^4m_\beta^2m_i^4m_j^2\Im[U_{i\alpha}U_{j\beta}U_{i\beta}^\ast U_{j\alpha}^\ast]~,\label{eq:Trnue3}
\end{align}
where $V_{\alpha i}$ and $U_{i\alpha}$ denote CKM and PMNS matrix, respectively.
By using Jarlskog invariants $J_{\mathrm{CP}}^{q}$ and $J_{\mathrm{CP}}^{l}$ defined as \cite{Tanabashi:2018oca}:
\begin{align}
	\Im [V_{i j} V_{k l} V_{i l}^{\ast} V_{k j}^{\ast}]
	&=
	J_{\mathrm{CP}}^q \sum_{m, n} \varepsilon_{i k m} \varepsilon_{j l n}~,\\
	\Im [U_{k \alpha} U_{l \beta} U_{k \beta}^{\ast} U_{l \alpha}^{\ast}]
	&=
	J_{\mathrm{CP}}^l \sum_{m, n} \varepsilon_{k l m} \varepsilon_{\alpha\beta n}~,
\end{align}
where $\varepsilon_{ijk}$ is completely antisymmetric tensor, we obtain
\begin{align}
	\tr\left([H_U,H_D]^3\right)
	&=
	6i\Delta_u \Delta_d J_{\mathrm{CP}}^q~,\label{eq:tr_JCPq}\\
	\tr\left([H_\nu,H_E]^3\right)
	&=
	-6i\Delta_\nu \Delta_e J_{\mathrm{CP}}^l~.\label{eq:tr_JCPl}
\end{align}
Here we have defined
\begin{align}
	\Delta_u
	&\equiv
	(m_u^2 - m_c^2)(m_c^2 - m_t^2)(m_u^2 - m_t^2)~,\\
	\Delta_d
	&\equiv
	(m_d^2 - m_s^2)(m_s^2 - m_b^2)(m_d^2 - m_b^2)~,\\
	\Delta_\nu
	&\equiv
	(m_1^2 - m_2^2)(m_2^2 - m_3^2)(m_1^2 - m_3^2)~,\\
	\Delta_e
	&\equiv
	(m_e^2 - m_{\mu}^2)(m_{\mu}^2 - m_{\tau}^2)(m_e^2 - m_{\tau}^2)~.
\end{align}
We can see from the Eqs.\eqref{eq:tr_JCPq} and \eqref{eq:tr_JCPl} that the sign of Jarlskog invariant for leptons is opposite to that for quarks
if the signs of $\tr\left([H_\nu,H_E]^3\right)$ and $\tr\left([H_U,H_D]^3\right)$ are the same.

If $M_U$ and $M_\nu$ are diagonal matrices, Eqs.\eqref{eq:Trud3} and \eqref{eq:Trnue3} can be rewritten as follows:
\begin{align}
	\tr\left([H_U,H_D]^3\right)
	&=
	6i \Delta_u \Im[(H_D)_{13}(H_D)_{32}(H_D)_{21}]~,\\
	\tr\left([H_\nu,H_E]^3\right)
	&=
	6i\Delta_{\nu}
	\Im[(H_E)_{13}(H_E)_{32}(H_E)_{21}]~.
\end{align}
The imaginary part $\Im[(H_{D,E})_{13}(H_{D,E})_{32}(H_{D,E})_{21}]$ can be rewritten in terms of the elements of mass matrices $M_D$ and $M_E$:

\begin{align}
	\Im [(H_i)_{13}(H_i)_{32}(H_i)_{21}]
	&=
	\{ | (M_i)_{11} |^2 - | (M_i)_{12} |^2 \} \Im [(M_i)_{21}
	(M_i^{\ast})_{22} (M_i^{\ast})_{31} (M_i)_{32}]\nonumber\\
	&\phantom{=}
	+ \{ | (M_i)_{11} |^2 - | (M_i)_{13} |^2 \} \Im [(M_i)_{21}
	(M_i^{\ast})_{23} (M_i^{\ast})_{31} (M_i)_{33}]\nonumber\\
	&\phantom{=}
	+ \{ | (M_i)_{32} |^2 - | (M_i)_{31} |^2 \} \Im [(M_i^{\ast})_{11}
	(M_i)_{12} (M_i)_{21} (M_i^{\ast})_{22}]\nonumber\\
	&\phantom{=}
	+ \{ | (M_i)_{21} |^2 - | (M_i)_{22} |^2 \} \Im [(M_i^{\ast})_{11}
	(M_i)_{12} (M_i)_{31} (M_i^{\ast})_{32}]\nonumber\\
	&\phantom{=}
	+ \{ | (M_i)_{33} |^2 - | (M_i)_{31} |^2 \} \Im [(M_i^{\ast})_{11}
	(M_i)_{13} (M_i)_{21} (M_i^{\ast})_{23}]\nonumber\\
	&\phantom{=}
	+ \{ | (M_i)_{21} |^2 - | (M_i)_{23} |^2 \} \Im [(M_i^{\ast})_{11}
	(M_i)_{13} (M_i)_{31} (M_i^{\ast})_{33}]\nonumber\\
	&\phantom{=}
	+ \{ | (M_i)_{12} |^2 - | (M_i)_{13} |^2 \} \Im [(M_i)_{22}
	(M_i^{\ast})_{23} (M_i^{\ast})_{32} (M_i)_{33}]\nonumber\\
	&\phantom{=}
	+ \{ | (M_i)_{33} |^2 - | (M_i)_{32} |^2 \} \Im [(M_i^{\ast})_{12}
	(M_i)_{13} (M_i)_{22} (M_i^{\ast})_{23}]\nonumber\\
	&\phantom{=}
	+ \{ | (M_i)_{22} |^2 - | (M_i)_{23} |^2 \} \Im [(M_i^{\ast})_{12}
	(M_i)_{13} (M_i)_{32} (M_i^{\ast})_{33}]\nonumber\\
	&\phantom{=}
	+ \Im [(M_i^{\ast})_{11} (M_i)_{12} (M_i)_{21} (M_i^{\ast})_{23}
	(M_i^{\ast})_{32} (M_i)_{33}]\nonumber\\
	&\phantom{=}
	+ \Im [(M_i^{\ast})_{11} (M_i)_{13} (M_i)_{21} (M_i^{\ast})_{22}
	(M_i)_{32} (M_i^{\ast})_{33}]\nonumber\\
	&\phantom{=}
	+ \Im [(M_i)_{11} (M_i^{\ast})_{12} (M_i)_{22} (M_i^{\ast})_{23}
	(M_i^{\ast})_{31} (M_i)_{33}]\nonumber\\
	&\phantom{=}
	+ \Im [(M_i^{\ast})_{12} (M_i)_{13} (M_i^{\ast})_{21} (M_i)_{22}
	(M_i)_{31} (M_i^{\ast})_{33}]\nonumber\\
	&\phantom{=}
	+ \Im [(M_i)_{11} (M_i^{\ast})_{13} (M_i^{\ast})_{22} (M_i)_{23}
	(M_i^{\ast})_{31} (M_i)_{32}]\nonumber\\
	&\phantom{=}
	+ \Im [(M_i)_{12} (M_i^{\ast})_{13} (M_i^{\ast})_{21} (M_i)_{23}
	(M_i)_{31} (M_i^{\ast})_{32}]~.\label{eq:HHH}
\end{align}


\end{document}